\documentclass[a4paper,12pt]{amsart}
\usepackage{graphicx}
\usepackage{psfrag}
\usepackage{amssymb}
\usepackage[colorlinks=true]{hyperref}
\usepackage{kbordermatrix}
\title{Littlewood--Richardson coefficients and integrable tilings}
\author[P.~Zinn-Justin]{Paul~Zinn-Justin}
\address{Paul Zinn-Justin, LPTMS (CNRS, UMR 8626), Univ Paris-Sud, 91405 Orsay Cedex, France; and LPTHE (CNRS, UMR 7589), Univ Pierre et Marie Curie-Paris6, 75252 Paris Cedex, France.}
\email{pzinn\,@\,lpthe.jussieu.fr}
\thanks{The author wants to thank 
J.~de Gier and B.~Nienhuis for providing him with their unpublished work,
and A.~Knutson and R.~Langer for encouragement and useful comments.}
\thanks{PZJ was supported by
EU networks
``ENRAGE'' MRTN-CT-2004-005616, ``ENIGMA'' MRT-CT-2004-5652,
ESF program ``MISGAM''
and ANR program ``GIMP'' ANR-05-BLAN-0029-01.}
\numberwithin{equation}{section}
\newtheorem{prop}{Proposition} 
\newtheorem{lemma}{Lemma}

\newtheorem{theorem}{Theorem}
\newtheorem*{corol}{Corollary}
\newcommand\e[1]{\ifinner{e^{#1}}\else{e^{\textstyle#1}}\fi}
\newcommand\oh{\frac{1}{2}}
\newcommand\F{\mathcal{F}}
\newcommand\G{\mathcal{G}}

\newcommand\Gf{\mathcal{G}_{\rm free}}
\renewcommand\S{{\bf S}}

\newcommand\R{{\bf R}}
\newcommand\T{{\bf T}}
\newcommand\Tp{{\bf \tilde T}_+}
\newcommand\Tm{{\bf \tilde T}_-}
\newcommand\Tpm{{\bf \tilde T}_{\pm}}
\newcommand\Tf{{\bf T}_{\rm free}}
\newcommand\Tff{\overline{\bf T}_{\rm free}}

\newcommand\ket[1]{\left|#1\right\rangle}
\newcommand\bra[1]{\left\langle#1\right|}
\newcommand\concat{\sqcup}

\newsavebox{\mybox}
\newcommand\myincludegraphics[3]{\sbox{\mybox}{\includegraphics[scale=#1]{#3}}
\includegraphics[width=#2\wd\mybox,height=\ht\mybox]{#3}}

\newdimen{\cellsize}
\newcommand\medboxes{\setlength{\cellsize}{14pt}\def\boxformat{}}
\newcommand\smallboxes{\setlength{\cellsize}{7pt}\def\boxformat{\scriptstyle}}
\def\boxformat{}
\newsavebox{\cellcontent}
\def\hidehrule#1#2{\kern-#1
  \hrule height#1 depth#2 \kern-#2 }%
\def\hidevrule#1#2{\kern-#1{\dimen\cellcontent=#1%
    \advance\dimen\cellcontent by#2\vrule width\dimen\cellcontent}\kern-#2 }%
\def\makeblankbox#1#2{\hbox{\lower\dp\cellcontent\vbox{\hidehrule{#1}{#2}%
    \kern-#1 
    \hbox to \wd\cellcontent{\hidevrule{#1}{#2}%
      \raise\ht\cellcontent\vbox to #1{}
      \lower\dp\cellcontent\vtop to #1{}
      \hfil\hidevrule{#2}{#1}}%
    \kern-#1\hidehrule{#2}{#1}}}}
\newcommand\cellify[1]{\defaultcell%
\sbox{\cellcontent}{\vbox to \cellsize{%
\vfill%
\hbox to \cellsize{\hfill$\boxformat #1$\hfill}
\vfill}}%
\rlap{\drawnbox}
\usebox{\cellcontent}}
\newcommand\tableau[1]{\vtop{\let\\\cr
\baselineskip -16000pt \lineskiplimit 16000pt \lineskip 0pt
\ialign{&\cellify{##}\cr#1\crcr}}}
\newcommand\defaultcell{\gdef\drawnbox{
\makeblankbox{0.2pt}{0.2pt}
}}
\newcommand\vdotscell{\gdef\drawnbox{\kern-1.6pt\vbox{\baselineskip=4pt\lineskiplimit=0pt\hbox{}\hbox{.}\hbox{.}\hbox{.}\hbox{}}}}
\newcommand\hdotscell{\gdef\drawnbox{\vbox to \cellsize{\hbox{\kern1pt$\ldotp\ldotp\ldotp$}}}}
\newcommand\vhdotscell{\gdef\drawnbox{\rlap{\kern-1.6pt\vbox{\baselineskip=4pt\lineskiplimit=0pt\hbox{}\hbox{.}\hbox{.}\hbox{.}\hbox{}}}\vbox to \cellsize{\hbox{\kern1pt$\ldotp\ldotp\ldotp$}}}}
\begin{document}
\begin{abstract}\vskip-0.5cm
We provide direct proofs of product and coproduct formulae for Schur functions where the
coefficients (Littlewood--Richardson coefficients) are defined as counting puzzles. The product
formula includes a second alphabet for the Schur functions, 
allowing in particular to recover formulae of [Molev--Sagan '99] and [Knutson--Tao '03] for factorial Schur functions.
The method is based on the quantum integrability of the underlying tiling model.
\end{abstract}
\maketitle\vskip-1.5cm\ 
{\small\baselineskip=9pt\lineskip=0pt%
\tableofcontents}\goodbreak

\section{Introduction}\label{intro}
Littlewood--Richardson coefficients are important integers related to Schur functions or equivalently
to the representation theory of the general linear group; they also appear in the cohomology of Gra\ss mannians.
Interesting combinatorial formulae for them have been given in \cite{KTW,KT}:
these coefficients count puzzles, i.e.\ certain tilings of a triangle where the tiles are decorated 
elementary triangles and rhombi. However all the proofs of this formula are fairly 
indirect e.g.\ rely on induction.

From the mathematical physicist's point of view, Littlewood--Richardson coefficients provide
a challenge. It is well-known \cite{JM-ff} that Schur functions are related to (two-dimensional)
{\em free fermions}. However it is not clear how to define Littlewood--Richardson coefficients in this framework.
In fact the contruction of \cite{KT}, as will be discussed here, suggests that the right way to describes them
involves {\em interacting}\/ fermions. Interestingly, the physical model in question has in fact
been studied in the physics literature. As pointed out in \cite{Pur}, it is equivalent to a model
of {\em random tilings}, the square-triangle triangle model,
which has been the subject of a lot of activity \cite{WidM,Kal,dGN-sqtri}.
Note that this equivalence is not particularly useful -- in order to solve the
square-triangle tiling model, one usually goes back to the model
of tilings of decorated triangles and rhombi.
The most important feature of this model for our purposes is that it is {\em integrable}:
the scattering of the elementary degrees of freedom (the aforementioned interacting fermions) is
factorized and thus satisfies the Yang--Baxter equation (with spectral parameters).

A general consequence of integrability is the existence of a commuting family of transfer matrices
that contains the original transfer matrix describing the discrete time evolution of the system.
In the case of free fermions, these transfer matrices precisely ``grow'' Schur functions. 
Here we find in fact two families of commuting matrices \cite{dGN-sqtriunpub}.
It is quite satisfying that computing their matrix elements
naturally produces one of the expressions that define Littlewood--Richardson coefficients, namely the
{\em coproduct}\/ formula. 
We thus obtain a direct proof of the combinatorial formula for them.

Furthermore, the integrability strongly suggests to introduce arbitrary spectral parameters into the model:
this corresponds to extending the original tiling model to a more general inhomogeneous model. This naturally
produces generalizations of the Littlewood--Richardson coefficients which are polynomials of the inhomogeneities. We recover this way several known formulae \cite{KT,MS-doubleschur} as well as a new one for the coefficients
in the expansion of the {\em product}\/ of factorial (or double) Schur functions. 

The paper is organized as follows. Section \ref{tilingmodel} is a presentation of the tiling model
which will be used throughout this paper.
Section \ref{focktm} presents the main ingredients in the
derivation of the coproduct formula: the Fock space/transfer matrix formalism.
Section \ref{mainproof} discusses the integrability of the tiling model and provides the proof
of the main theorem used in section \ref{nonequiv}, which is the derivation of the coproduct formula.
Finally, section \ref{equiv} describes the inhomogeneous model and its application to 
product formulae for factorial Schur functions. The appendix briefly discusses the equivalence to a
square-triangle-rhombus tiling model.

\section{The tiling model}\label{tilingmodel}
We provide here our own formulation of the model of tilings by decorated rhombi and triangles which is the basis
of \cite{KT} (we actually extend it slightly by defining two additional tiles).
\subsection{Tiles}
The model is defined by filling a domain contained inside the triangular lattice with tiles of the form
shown on Fig.~\ref{figtiles}. More precisely, one can use {\em either}\/ the colored lines inside the tiles {\em or}\/
the symbols on the edges to decide if adjacent tiles match: that is, symbols must coincide on the edges, or
equivalently green and red lines must propagate across edges. Colored lines or edge labels
can be thought of as two equivalent ways to encode the four possible states of 
edges, which each have their advantages. We shall mostly use colored lines in what follows.
The correspondence of edge labels with the notation of \cite{KT} is $-\leftrightarrow 0$,
$+\leftrightarrow 1$,
$0\leftrightarrow 10$, $\tilde 0\leftrightarrow 01$. The shading of the tiles $\gamma$ will be explained later (cf
a similar shading in \cite{KT}).

\newcommand{\edgerelabelling}{\psfrag{0}{$\scriptscriptstyle\! -$}\psfrag{1}{$\scriptscriptstyle\! +$}\psfrag{10}{$\scriptscriptstyle\, 0$}\psfrag{01}{$\scriptscriptstyle\, \tilde 0$}}
\begin{figure}
\edgerelabelling
\psfrag{a}{$\alpha_-$}\psfrag{b}{$\alpha_+$}\psfrag{c}{$\beta_-$}\psfrag{d}{$\beta_+$}\psfrag{e}{$\beta_0$}\psfrag{f}{$\gamma_-$}\psfrag{g}{$\gamma_+$}\psfrag{h}{$\gamma_0$}
\myincludegraphics{0.5}{1.15}{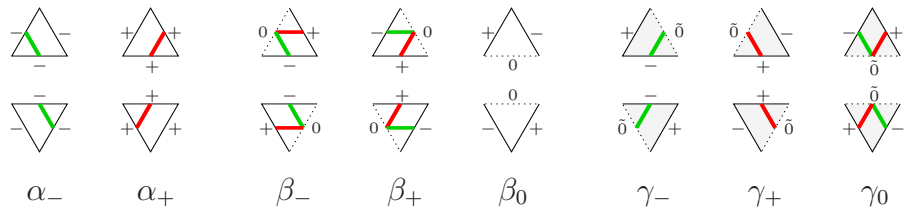}%
\caption{The tiles.}\label{figtiles}
\end{figure}

Note that we have classified tiles in pairs; this is because the two tiles of type $\beta_-$, $\beta_+$, $\beta_0$, $\gamma_-$, $\gamma_+$, $\gamma_0$
always appear together on adjacent triangles to form rhombi, which we have illustrated by using dotted lines. There is some freedom however in how to reconnect
the two tiles of type $\alpha_\pm$: an upper tile of type $\alpha_-$ can have on its left any number
of pairs of tiles of type $\beta_+$; 
the only way this series can end is with a lower tile of type $\alpha_-$. Similarly for the tiles of type $\alpha_+$,
with series of tiles of type $\beta_-$ on its right.
In what follows it will be convenient to consider tilings of the whole plane.
However, we shall see that with our ``boundary conditions'' (conditions at left and right infinity),
all such series of tiles of type $\beta_\pm$ will be necessarily finite. 
\begin{figure}
\myincludegraphics{0.5}{1.15}{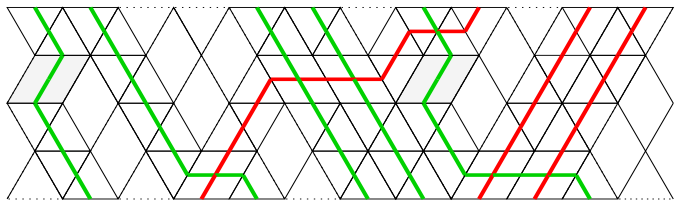}
\caption{An example of tiling, using tiles $\alpha_\pm$, $\beta_{\pm0}$, $\gamma_-$.}
\end{figure}

\subsection{Paths}\label{secpaths}
At this stage we can forget about the underlying triangles and rhombi and simply keep track of the paths
formed by the green and red lines. Consider a horizontal line in the triangular lattice.
Each edge can be in only three states: empty or occupied by a green or red line. In what follows
we shall number edges using alternatingly half-odd-integers and integers to take into account the
nature of the lattice. (One could of course get rid of this issue by applying an additional shift
by a half-step say to the right, but that would break the left-right symmetry of the model, and we do not
choose to do so here.)

\begin{figure}
\myincludegraphics{0.5}{1.15}{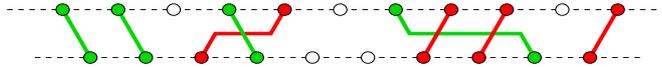}
\caption{An example of evolution of paths from one time step to the next.}\label{figpaths}
\end{figure}

We now analyze what happens to the lines during one ``time step'',
that is as one moves (upwards) from one horizontal line to the next.
Let us first ignore the tiles $\gamma_{0,-,+}$. This corresponds to the original tiling model of \cite{WidM} which also
occurs in \cite{KT} (in the non-equivariant case). 
Then the rules are as follows: a green (resp.\ red) line moves one half-step to the
right (resp.\ left) if there is currently no particle of the opposite kind at this spot.
The only situation left to consider is when lines of opposite colors are adjacent, with the green line at the left.
Then two scenarios occur: either the green line crosses all the red lines at its right until it finds an empty
spot, or the red line crosses all the green lines at its left until it finds an empty spot.
This is shown on Fig.~\ref{figpaths}.

If we add the tiles $\gamma_-$ (resp.\ $\gamma_+$), then green (resp.\ red) lines have the additional possibility of moving
in the opposite direction as normally, on condition that the spot is free. If we add the tile $\gamma_0$, then
green and red lines are allowed to simply cross each other as if they did not see each other.

\section{Fock spaces and transfer matrices}\label{focktm}
We first define the notion of Fock space. The idea is that
to encode the possible configurations of tilings on a given horizontal line into a Hilbert space. 
But first we describe another Fcok space which is slightly simpler 
(only two states per site instead of three) and will play an important role.
\subsection{Fermionic Fock space $\F$}\label{fockfree}
The Fock space $\F$ is an infinite dimensional Hilbert space with canonical orthonormal basis defined as follows.
Each element $\ket{f}$ of the basis 
is indexed by a map $f$ from $\mathbb{Z}+\oh$ to $\{-1,1\}$ such that
there exists $N_-$, $N_+$ such that $f(i)=-1$ for $i<N_-$ and $f(i)=+1$ for $i>N_+$. 
Call $N_+(f)$ (resp.\ $N_-(f)$) the smallest (resp.\ largest) such integer.

We shall represent the $-1$'s (resp.\ $1$'s) as green (resp.\ red) particles or dots.

There is a notion of ``charge'' which can be thought of as follows: each green particle has charge $-1$ and each
red particle 
has charge $+1$. This is ill-defined because there is an infinite number of particles, so we need
a reference state. Define $\ket{\varnothing}$ (the vacuum state) 
to be the state such that there are only green particles to the left of zero and only red particles to the right. 
The corresponding map
from $\mathbb{Z}+\oh$ to $\{ -1,1\}$ is the sign map. $\ket{\varnothing}$ has by definition zero charge.
This way, the charge of any state $\ket{f}$ is given by $c(f):=\sum_{i\in\mathbb{Z}+\oh}(f(i)-\text{sign}(i))$.
The charge is always an even number (we use twice the standard convention, for reasons that will become
clear).

Define the shift operator $\S$: it is defined by $\S\ket{f}=\ket{f'}$ with $f'(i+1)=f(i)$ for all $i$.
$\S$ decreases the charge by $2$, and is an isomorphism between subspaces of different charge.

In a subspace of given charge, basis elements can alternatively be indexed by
Young diagrams \cite{JM-ff} (see also \cite{PZJ-houches}). 
The correspondence goes as follows. Rotate the Young diagram 45 degrees,
assign green dots and red dots to edges of either orientation as indicated on the picture:
\[
\includegraphics[width=7cm]{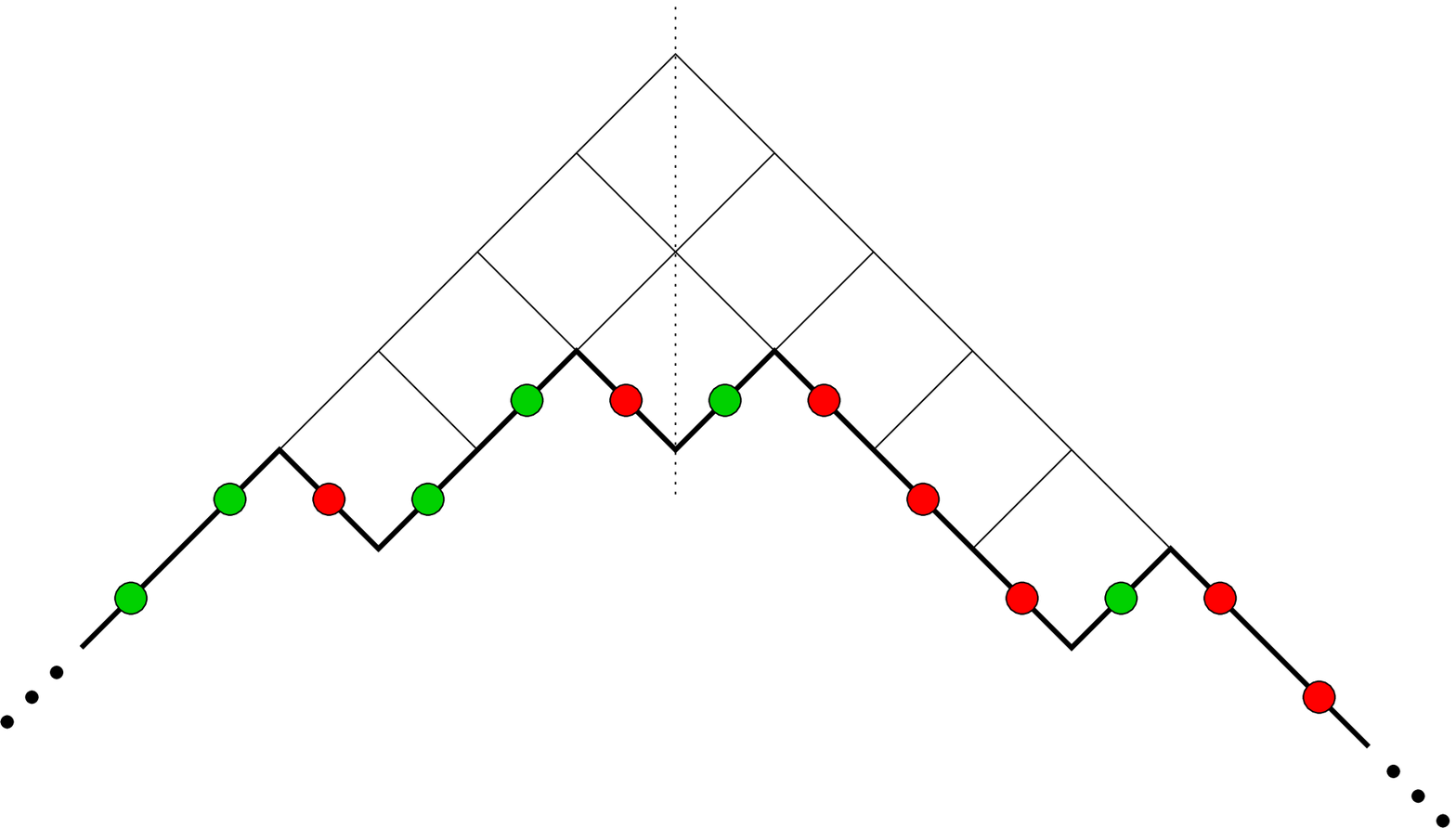}
\]
One can then flatten the line and obtain a configuration of green and red dots. There remains the arbitrariness
in shifting the line, or equivalently in the charge.
Here we shall only consider the case of zero charge, in which case the convention is that the diagonal line
(dotted line on the picture) represents the zero. In such a way to any Young diagram $\lambda$ we associate
a state simply denoted by $\ket{\lambda}$, and all the basis vectors of the subspace with zero charge are
recovered this way. In the case of the empty diagram we recover our vacuum state
$\ket{\varnothing}$.

Finally, define for future use $\F_{+,k}$ (resp.\ $\F_{-,k}$) to be the span of the $\ket{f}$ such that $N_+(f)\le k$
(resp.\ $N_-(f)\ge -k$).

\subsection{Fock space $\G$ of the tiling model}
The Fock space $\G$ can be similarly described as follows.
Basis vectors $\ket{f}$ of $\G$
are indexed by maps $f$ from $\mathbb{Z}+\oh$ to $\{-1,0,+1\}$
such that there exists $N_-$, $N_+$ such that $f(i)=-1$ for $i<N_-$ and $f(i)=+1$ for $i>N_+$. 

The correspondence to configurations of the tiling model described in the previous section is as follows:
each basis vector of $\G$ encodes a horizontal line in a configuration of tiles; thus,
$-1\equiv -$ correspond to a green particle, $0$ to an empty spot and $+1\equiv +$ to a red particle. 
Since the model is translationally invariant the choice of an origin is irrelevant;
however, note that successive lines have all sites shifted by half a step, which means that $\G$ can only
describe rows of a given parity, not both at the same time. We shall come back to this point below.

Define $N_\pm(f)$ similarly as before: that is, $N_-(f)$ is the location of the leftmost empty spot or red particle minus one half,
whereas $N_+(f)$ is the location of the rightmost empty spot or green particle plus one half.
We can also define two more numbers which will be useful:
$N_{-0}(f)$ is the location of the leftmost red particle minus one half, 
whereas $N_{+0}(f)$ is the location of the rightmost green particle plus one half.

There are two ``quantum numbers'' in $\G$. The first one, the charge, is defined in $\G$ as in
$\F$ by $c(f):=\sum_{i\in\mathbb{Z}+\oh}(f(i)-\text{sign}(i))$ i.e.\ green particles have charge $-1$,
red particles have charge $+1$, empty spots have zero charge. 
The charge is an integer with arbitrary parity.

The second quantum number, the ``emptiness number'', is simply the number of zeroes: $e(f):=\# \{i: f(i)=0\}$.

Intuitively, the conservation of the two quantum numbers is related to the conservation of the number of lines of either color in any finite region. Since we have an infinite system, particles can however ``leak to infinity'', which results in variation of these quantum numbers.

There is again a shift operator, denoted by $\S$, which to $\ket{f}\in\G$ associates $\ket{f'}$ 
such that $f'(i+1)=f(i)$ for all $i$.
$\S$ preserves the emptiness number, and decreases the charge by $2$.

\subsection{From $\F$ to $\G$}
There are two types of maps we need to define from $\F$ to $\G$.

There is the obvious inclusion map from $\F$ to $\G$. This identifies $\F$ with the subspace of $\G$
with zero emptiness number.


The less obvious map $\concat$
takes two basis elements $\ket{f_-}$ and $\ket{f_+}$ 
and produces $\ket{f}=\ket{f_-}\concat\ket{f_+}$ in $\G$ 
such that 
\[
f(i)=\begin{cases}
\oh(f_-(i)-1)&i<0\\
\oh(f_+(i)+1)&i>0
\end{cases}
\]
In other words it ``concatenates'' the two words by discarding the right of $f_-$ and the left of $f_+$.

\begin{figure}
\psfrag{0}{$0$}\psfrag{a}{$\concat$}\includegraphics[width=11cm]{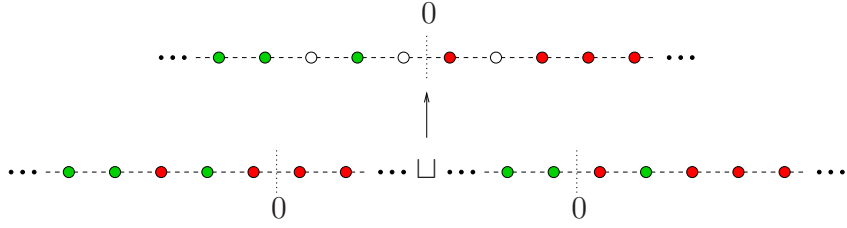}
\caption{The concatenation map.}
\end{figure}

More generally, define for $k\in\mathbb{Z}$
\[
\ket{f_-}\concat_k\ket{f_+}=
\S^{-k}\ket{f_-}\concat \S^{k}\ket{f_+}
\]
This map $\concat_k$ is injective if one restricts to $\ket{f_-}$ and $\ket{f_+}$ such that $N_+(f_-)\le k$ 
and $N_-(f_+)\ge-k$,
which is the only situation where we shall use $\concat_k$. 
We thus consider from now on $\concat_k$ as a linear map
from $\F_{+,k}\otimes\F_{-,k}$ to $\G$.

It is an easy calculation that if $\ket{f}=\ket{f_-}\concat_k\ket{f_+}$,
\begin{align*}
c(f)&=\frac{1}{2}(c(f_-)+c(f_+))\\
e(f)&=\frac{1}{2}(c(f_-)-c(f_+))+2k
\end{align*}

The image of $\concat_k$, denoted by $\Gf\subset\G$,
is exactly the span of the $\ket{f}$ such that $N_{-0}(f)\ge 0$ and $N_{+0}(f)\le 0$.


{\em Remark 1}: intuitively, this second operation has the following meaning. When the sets of green and red particles
are widely separated from each other (green ones being on the left and red ones on the right), then each of them
behaves like a system of fermions (the fermionic character being the condition that there can be at most
one particle per site).

Next we shall define transfer matrices. In fact, we should say a few words on
what we mean by a ``transfer matrix'' here because of the fact that we
are dealing with infinite-dimensional spaces. A transfer matrix is defined here as a matrix, that is
a collection of entries $(\T_{f,g})$ where $f$ and $g$ index the canonical basis of $\F$ or $\G$. 
It is tempting
to associate to it a linear operator $\T$ on $\F$ or $\G$, such that $\T_{f,g}=\bra{f}\T\ket{g}$,
but this is problematic because sometimes
the action of $\T$ leads to an infinite linear combination of basis elements,
which would require discussing the convergence of summations. However in all that follows,
whenever we have two transfer matrices $\T$ and $\T'$, the
product $(\T\T')_{f,h}=\sum_g \T^{}_{f,g}\T'_{g,h}$ only involves finite sums
and is therefore well-defined; so that we can safely ignore this subtlety
and manipulate transfer matrices as operators.

\subsection{The transfer matrix of free fermions}
We first define the usual dynamics for free fermions that leads to Schur functions, see e.g.\ \cite{PZJ-houches}.

\begin{figure}
\includegraphics[width=9cm]{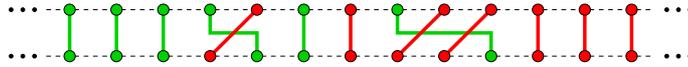}
\caption{An example of evolution of the free fermionic model.}\label{figff}
\end{figure}

The transfer matrix, denoted by $\Tf(u)$, is most simply described by considering red dots as lines propagating
(similarly as green and red lines in the tiling model). In this case the rule of evolution for red lines is
that at each step, they can either move straight upwards or upwards and one step to the right on condition that
no two lines touch each other. An example is given on Fig.~\ref{figff}. Furthermore, sufficiently far to the right, we
impose that red lines go straight upwards. This way any evolution only involves finitely many moves to the right:
we then assign a weight of $u$ to each such move. Explicitly, $\bra{f}\Tf(u)\ket{f'}$ equals the sum over
configurations of the form of Fig.~\ref{figff} where the initial configuration (at the bottom) is described by $f'$
and the final configuration (at the top) is described by $f$, of $u$ to the power the number of moves to the right.

One remark is in order. One can of course also assign lines to the green dots and formulate the rules in
terms of the green lines (see \cite{PZJ-houches} for details).
These lines have also been 
represented on Fig.~\ref{figff}. The rule is that at each time step green lines can move up half-way then
arbitrarily far to the left then up again, but in such a way that they do not touch any other green lines
along the way. The weight of $u$ is given to each crossing of green and red lines.

$\Tf(u)$ breaks the ``particle--hole'' symmetry that exchanges left and right, green and red lines,
since the rules are clearly different for the two types of lines.
One can therefore introduce a mirror-symmetric transfer matrix $\Tff(u)$. It is defined similarly as $\Tf(u)$,
but this time, the green lines are allowed to go either straight upwards or upwards and 
one step to the left. Each left move is given a weight of $u$.

Both $\Tf(u)$ and $\Tff(u)$ preserve the charge.
\goodbreak

Finally, we have the following important formulae:
\begin{lemma}\label{basicrel}
\[
s_{\lambda/\mu}(u_1,\ldots,u_n)=\bra{\mu}\prod_{i=1}^n \Tf(u_i)\ket{\lambda}
\]
is the skew Schur function associated with the Young diagram $\lambda$. For $\mu=\varnothing$,
\[
s_\lambda(u_1,\ldots,u_n)=\bra{\varnothing}\prod_{i=1}^n \Tf(u_i)\ket{\lambda}
\]
is the Schur function associated with the Young diagram $\lambda$.
Similarly,
\[
s_{\lambda^T/\mu^T}(u_1,\ldots,u_n)=\bra{\mu}\prod_{i=1}^n \Tff(u_i)\ket{\lambda}
\]
where $\lambda^T$ is the transpose of $\lambda$, and in particular
\[
s_{\lambda^T}(u_1,\ldots,u_n)=\bra{\varnothing}\prod_{i=1}^n \Tff(u_i)\ket{\lambda}
\]
The most general formula is
\[
s_{\lambda/\mu}(u_1,\ldots,u_m/v_1,\ldots,v_n)=\bra{\mu}\prod_{i=1}^m \Tf(u_i)
\prod_{i=1}^n \Tff(v_i)\ket{\lambda}
\]
that is the supersymmetric skew Schur function, which leads for $\mu=\varnothing$ to the usual supersymmetric Schur function
\[
s_\lambda(u_1,\ldots,u_m/v_1,\ldots,v_n)=\bra{\varnothing}\prod_{i=1}^m \Tf(u_i)
\prod_{i=1}^n \Tff(v_i)\ket{\lambda}
\]
\end{lemma}
{\em Remark 2}: as will be apparent in the proof,
the expressions in the lemma are independent of the ordering of the products,
which is why we left them unspecified.
This implies the commutation relations
\[
[\Tf(u),\Tf(v)]=0\qquad [\Tf(u),\Tff(v)]=0\qquad [\Tff(u),\Tff(v)]=0
\]
which are also a consequence of the more general results of the next section.
\begin{proof}
There are several simple proofs of this standard result. 
Note first that taking products of $n$ transfer matrices amounts to stacking together $n$ rows made of the
paths defined above.
One proof involves a bijection between these paths and the appropriate
tableaux that one uses to define supersymmetric Schur functions.
Another proof, which we sketch here, is 
to use the Lindstr\"om--Gessel--Viennot (LGV) formula \cite{Lind,GV}. We apply it 
to the green lines to the right of $N_-(\lambda)$ (those to the left
necessarily go straight). There are exactly $\lambda'_1$ such lines,
where $\lambda'_1$ is the number of non-zero rows of $\lambda$.
This leads us to compute the evolution for a single line from initial location $k$ to final location $k+j$.
Noting that the rules of evolution are translationally invariant, one can introduce a generating function
$h^{(i)}(x)=\sum_{j\ge0}h^{(i)}_j x^j$ for the time step corresponding to $\Tf(u_i)$ 
and $e^{(i)}(x)=\sum_{j\ge0}e^{(i)}_j x^j$
for the time step corresponding
to $\Tff(v_i)$. 
$h^{(i)}_j$ and $e^{(i)}_j$ are the numbers of ways to move $j$ steps to the left for a single green
line and a single time step, so we immediately find
\[
h^{(i)}(x)=(1-x u_i)^{-1}\qquad e^{(i)}(x)=1+xv_i
\]
Composing the transfer matrices amounts to multiplying the generating series, so we find the evolution
for a single green line to be given by the generating series
\[
h(x;u_1,\ldots,u_m/v_1,\ldots,v_n)=\frac{\prod_{i=1}^n(1+x v_i)}{\prod_{i=1}^m(1-x u_i)}
\]
which is exactly the generating series of 
the supersymmetric analogues of completely symmetric functions
i.e.\ Schur functions $h_j$
corresponding to one row:
\[
h(x;u_1,\ldots,u_m/v_1,\ldots,v_n)=\sum_{j\ge 0} h_j(u_1,\ldots,u_m/v_1,\ldots,v_n) x^j
\]
Applying the LGV formula produces the Jacobi--Trudi identity
for supersymmetric Schur functions
\[
s_{\lambda/\mu}(u_1,\ldots,u_m/v_1,\ldots,v_n)=
\det\left(h_{\lambda_j-\mu_i+i-j}(u_1,\ldots,u_m/v_1,\ldots,v_n)\right)_{1\le i,j\le \lambda'_1}
\]
\end{proof}

\subsection{The transfer matrix of the tiling model}\label{tmorig}
Before defining the transfer matrix of the tiling model, one must discuss
the problem of the numbering of successive rows.
Since the edges are shifted by a half-step during one time unit, there are two symmetric choices:
shift everything one half-step to either the right or the left (either before or after the evolution,
since it is translationally invariant).
The resulting transfer matrices are called respectively
$\T_+$ and $\T_-$.
$\bra{f}\T_\pm\ket{g}$ is then 
defined as the number of possible ways paths can move from the initial configuration encoded
by $g$ to the final configuration encoded by $f$ in one time step according to the rules of evolution
described in section \ref{secpaths}, using only
the $2\times 5$ tiles $\alpha$ and $\beta$ of Fig.~\ref{figtiles}. In principle one could assign weights to
the different tiles but we shall not need to do so here. Note the relation $\T_+=\S \T_-$. 
In the case of a two-row evolution, 
one can introduce $\T^2:=\T_+\T_-=\T_-\T_+$, which has the advantage that it restores the left-right symmetry.\footnote{Despite the notation,
$\T^2$ is not the square of an operator on $\G$; if one insisted that it be so,
one would have to define $\T$ as acting on $\G\oplus\G$,
with $\T=\left(\begin{smallmatrix}0&\T_+\\ \T_-&0\end{smallmatrix}\right)$.}

Note that sufficiently far to the left, there are only
green lines and these necessarily move one half-step to the left. On the contrary, far to the right,
one has red lines that move one half-step to the right.
This observation allows us to conclude that $\T_\pm$ changes the charge by $\mp 1$ and increases the emptiness
number by $1$,
or equivalently $\T^2$ preserves the charge, but increases the emptiness number by $2$.

We now prove some additional properties of $\T^2$.
\begin{lemma}\label{motion}For any pair of basis states $f$ and $g$,
\[
\bra{f}\T^2\ket{g}\ne 0\ \Rightarrow\ N_{-0}(f)\ge N_{-0}(g)+1\ \text{and}\ N_{+0}(f)\le N_{+0}(g)-1
\]
\end{lemma}
\begin{proof}Since particles of the same color never cross, the leftmost or rightmost particles remain the same
during time evolution. The lemma then follows from the fact that red (resp.\ green) particles move at least one
half-step to the right (resp.\ left) at each time step.
\end{proof}

\begin{lemma}\label{freeaction}
If $\ket{f}=\ket{f_-}\concat\ket{f_+}\in\Gf$, with $N_+(f_-)\le 0$ and $N_-(f_+)\ge0$,
then $\T^2\ket{f}\in\Gf$, and
\[
\T^{2k} \ket{f}=\ket{f_-}\concat_k\ket{f_+}
\qquad
\forall k\in\mathbb{Z}_+
\]
\end{lemma}
\begin{proof}
Such a state $\ket{f}$ has the properties that all green lines are to the left of
red lines, so no crossings ever occur. Therefore,
the green (resp.\ red) lines move one half-step to the left
(resp.\ right) at each time step. This is all that the lemma says.
\end{proof}

\begin{lemma}\label{freeactionb}Call $p=\max(N_{+0}(f),-N_{-0}(f))$.
Then $\T^{2p}\ket{f}\in \Gf$ and
there exist unique coefficients $c^{f}_{g,h}$ such that
\[
\T^{2k} \ket{f}=\sum_{g,h} c^{f}_{g,h} \ket{g}\concat_k\ket{h}
\qquad
\forall k\ge p
\]
\end{lemma}
Intuitively, this lemma says that no matter what the initial state is, eventually all possible crossings
will take place and we shall be left with a linear combination of states which are all such that all green
lines are the to the left of red lines.
\begin{proof}
This is combination of the two preceding lemmas.
Let $\ket{f}$ be a basis state, and set $p=\max(N_{+0}(f),-N_{-0}(f))$.
Then according to lemma \ref{motion}, $T^{2p}\ket{f}$ is a linear combination
of basis states $\ket{g}$ such that $N_{-0}(g)\ge 0$ and $N_{+0}(g)\le 0$, i.e.\ 
$\T^{2p}\ket{f}\in\Gf$.

By definition of $\Gf$, this implies that there exist uniquely defined coefficients
$c^{f}_{g,h}$ such that 
$\T^{2p} \ket{f}=\sum_{g,h} c^{f}_{g,h} \ket{g}\concat_p\ket{h}$,
where the summation is restricted
to couples $(g,h)$ such that
$N_+(g)<p$, $N_-(h)>-p$.
Set all other entries $c^{f}_{g,h}$ to zero.
The formula of the lemma then follows by application of lemma \ref{freeaction}.
\end{proof}
\begin{corol}
If $\ket{\lambda}\in\F\subset\G$ is a state with zero charge and zero emptiness number 
and $p=\max(N_+(f),-N_-(f))$, then
there exist unique coefficients $c^{\lambda}_{\mu,\nu}$ such that
\[
\T^{2k} \ket{\lambda}=\sum_{\mu,\nu} c^{\lambda}_{\mu,\nu} \ket{\mu}\concat_k\ket{\nu}
\qquad
\forall k\ge p
\]
where the sum is over pairs of Young diagrams.
\end{corol}
\begin{proof}
Such a state $\ket{\lambda}$ has zero charge and zero emptiness number. The formula of lemma \ref{freeaction}
implies that all pairs $(g,h)$ that contribute to the sum satisfy $0=c(\lambda)=\oh(c(g)+c(h))$
and $0=e(\lambda)=\oh(c(g)-c(h))$, so that they have zero charge themselves. They can therefore
be indexed by Young diagrams.
\end{proof}

\subsection{The two families of commuting transfer matrices}\label{tmcommut}
We define here two one-parameter families of transfer matrices $\Tpm(u)$, which are closely connected to $\T_\pm$.

Naively, one would like to define $\Tpm$ as transfer matrices which, in addition to the usual tiles $\alpha$
and $\beta$, allow one more type of tiles $\gamma_\pm$, but at a cost of a certain weight $x$.
However, this turns out to be inconvenient for the boundary conditions we have in mind, so we use the following
definition instead, using the inverse $u=x^{-1}$ of the natural parameter $x$ (which will reoccur in section
\ref{mainproof}).
$\Tpm(u)$ is defined similarly as $\T_\pm$, but with three new ingredients: (i) we allow the extra tiles of
type $\gamma_\pm$ 
and (ii) we give a weight of $u$ to each pair of tiles of type $\alpha_\pm$, and to each
pair of tiles of type $\beta_\pm$ (all the signs being the same as that of the transfer matrix).
(iii) we impose the following condition at infinity:
sufficiently far to the left and to the right, the effect of $\Tm(u)$ (resp.\ $\Tp(u)$) 
must be to push either type of lines
one half-step to the right (resp.\ left). Taking into account that this half-step is absorbed
into the definition of the matrix, we find that
at infinity $\Tpm(u)$ behaves like the identity. 
This ensures that only a finite number of tiles of type $\alpha_\pm$ or $\beta_\pm$ ever
occur in the evolution w.r.t.\ $\Tpm(u)$. Thus each entry of $\Tpm(u)$ is a polynomial in $u$.

The boundary conditions also imply that $\Tpm(u)$ preserves
the charge and the emptiness number.

We now list some properties of $\Tpm(u)$.
\begin{lemma}\label{motionb}
For any pair of basis states $f$ and $g$,
\begin{gather*}
\bra{f}\Tp(u)\ket{g}\ne 0\ \Rightarrow\ N_{-0}(f)\ge N_{-0}(g)-1\ \text{and}\ N_{+0}(f)\le N_{+0}(g)\\
\bra{f}\Tm(u)\ket{g}\ne 0\ \Rightarrow\ N_{-0}(f)\ge N_{-0}(g)\ \text{and}\ N_{+0}(f)\le N_{+0}(g)+1
\end{gather*}
\end{lemma}
\begin{proof}
Same proof as lemma \ref{motion}, but this time green particles can move one half-step to the right for $\Tm$ and red particles
can move one half-step to the left for $\Tp$.
\end{proof}

\begin{lemma}\label{freeactionc}
$\ket{f}=\ket{f_-}\concat\ket{f_+}\in\Gf$, then
\begin{align*}
\Tm(u)\ket{f}&=\Tff(u)\ket{f_-}\concat\ket{f_+}\\
\Tp(u)\ket{f}&=\ket{f_-}\concat\Tf(u)\ket{f_+}
\end{align*}
\end{lemma}
\begin{proof}By inspection. If there are only green particles to the left of 0 and red particles to the right of
0, no crossings can occur. For $\Tm(u)$, this leaves two possibilities for green particles: going left one half-step
(pair of tiles $\alpha_-$) 
with a weight of $u$ or going right one half-step (pair of tiles $\gamma_-$) with a weight of $1$;
and only one possibility for red particles, going right 
one-half step (pair of tiles $\alpha_+$) with a weight of $1$.
Adjusting the locations by shifting everything one half-step to the left, we obtain exactly
$\Tff(u)$ for green particles and no evolution for red particles. The reasoning is the same for
$\Tp(u)$.
\end{proof}
Note the similarity with lemma \ref{freeaction}. 
However, an important difference with the situation of lemma \ref{freeaction} is that it is not true
that $\ket{f}\in\Gf$ implies $\Tpm(u)\ket{f}\in\Gf$, so in general
one cannot iterate the argument.

{\em Remark 3} (followup of {\em Remark 1}):
intuitively, this means that when the sets of green and red particles
are widely separated from each other then our transfer matrices make them evolve like free fermions.
When green and red particles mix, it is not the case any more: 
in this sense what we have is a system
of two species of fermions interacting with each other.

\begin{lemma}\label{crossaction}
The transfer matrices $\Tpm(u)$ leave $\F\subset\G$ stable, and:
\begin{align*}
\Tm(u)\ket{f}&=\Tff(u)\ket{f}& \forall \ket{f}\in \F\\
\Tp(u)\ket{f}&=\Tf(u)\ket{f}& \forall \ket{f}\in \F
\end{align*}
\end{lemma}
\begin{proof}
$\F$ viewed as a subspace of $\G$ is simply the subspace of zero emptiness number, and the $\Tpm(u)$ preserve
this number, so they leave $\F$ stable.

The rest of the reasoning is again by inspection. Consider the action of $\Tp(u)$ on a state with no empty
spots. Remembering that the final state should also have no empty spots (this is in fact a consequence
of the boundary conditions, as already explained),
we conclude that tiles of type $\alpha_+$ are forbidden (consider the leftmost tile of type upper $\alpha_+$;
the only allowed tile left of it
is the empty tile $\beta_0$). Similarly tiles of type $\beta_-$ are forbidden (sequences of such tiles
always end with tiles of type $\alpha_+$).
All that we are left with is half-steps to the left (tiles $\alpha_-$ and $\gamma_+$) and crossings of the
type $\beta_+$, that is one green line crossing a series of red lines.
But up to an overall half-step to the left, these crossings are exactly those that occur between red and green
lines in the free fermionic model, compare Figs.~\ref{figpaths} and \ref{figff}. The weight of $u$ is given to each
each pair of $\beta_+$ tiles, that is
to each crossing, which is the same weight that is given in the free fermionic model.

A similar reasoning works for $\Tm(u)$ and $\Tff(u)$.
\end{proof}

\begin{theorem}\label{mainthm} All transfer matrices $\Tm(u)$, $\Tp(v)$, $\T$ commute:
\[
[\tilde \T_{\epsilon}(u),\tilde\T_{\epsilon'}(v)]=0
\qquad
[\tilde\T_\epsilon(u),\T_{\epsilon'}]=0\qquad \epsilon,\epsilon'\in\{+,-\}
\]
\end{theorem}
This is the central result of this section, embodying the integrability of the model.
Its proof is the subject of next section.

\section{Yang--Baxter equation and proof of the commutation theorem}\label{mainproof}
We now discuss the integrability of the tilings model, that is the various forms of the Yang--Baxter
and how they imply the commutation relations of theorem \ref{mainthm}. 
\subsection{$R$-matrix and Yang--Baxter equation}
The first idea is to decompose the action of the transfer matrices as products of elementary blocks, called $R$-matrices.
It is convenient in this section to go back to the labels $+,-,0,\tilde{0}$ on the edges because using it,
the tiles are explicitly $\mathbb{Z}_3$-invariant.
We can then choose as elementary blocks two adjacent triangles. They have three possible orientations,
which correspond to three possible rhombi.

\begin{figure}
\psfrag{I}{$1$}\psfrag{x}{$x$}\edgerelabelling
\myincludegraphics{0.5}{1.15}{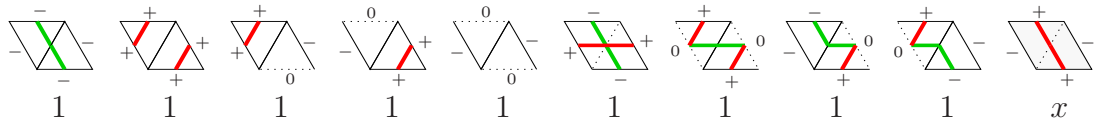}
\caption{The $R$-matrix in one of its three orientations.}\label{figLmat}
\end{figure}

Let us choose for example the orientation as shown on Fig.~\ref{figLmat}, and allow the tiles
$\alpha_\pm$, $\beta_{\pm0}$, $\gamma_+$,
with the weights indicated below. $x$ is the {\em spectral parameter}.
In fact the only rule is that pairs of shaded tiles get a weight of $x$.
The weights look different but are actually equivalent to the weights of $\Tp(v)$, with $v=1/x$ (this will be explained
in the next section). The colored lines are only decorative here and the focus is on the edge numbers.
We can encode this information into a $3\times3\times 3\times3$ tensor as follows.
$\R_{i,j,k,l}(x)$, where each index lives in $\{-,+,0\}$,
is the weight of the rhombus
with edges $i,j,k,l$ read counterclockwise starting from the left edge (or the right edge,
the tiles being 180 degrees symmetric).

$\R(x)$ is usally considered as a matrix by associating to it
the map from $\mathbb{C}^3\otimes \mathbb{C}^3$ (corresponding to the left and bottom edges)
to itself (corresponding to the right and top edges); explicitly,
\renewcommand{\kbldelim}{(}
\renewcommand{\kbrdelim}{)}
\renewcommand{\kbrowstyle}{\scriptscriptstyle}
\renewcommand{\kbcolstyle}{\scriptscriptstyle}
\begin{equation}
\R(x)=\kbordermatrix{
&-,-&-,+&-,0&+,-&+,+&+,0&0,-&0,+&0,0\\
-,-& 1 & 0 & 0 & 0 & 0 & 0 & 0 & 0 & 0 \\
-,+& 0 & x & 0 & 0 & 0 & 1 & 1 & 0 & 0 \\
-,0& 0 & 0 & 1 & 0 & 0 & 0 & 0 & 0 & 0 \\
+,-& 0 & 0 & 0 & 1 & 0 & 0 & 0 & 0 & 0 \\
+,+& 0 & 0 & 0 & 0 & 1 & 0 & 0 & 0 & 0 \\
+,0& 0 & 1 & 0 & 0 & 0 & 0 & 0 & 0 & 0 \\
0,-& 0 & 1 & 0 & 0 & 0 & 0 & 0 & 0 & 0 \\
0,+& 0 & 0 & 0 & 0 & 0 & 0 & 0 & 1 & 0 \\
0,0& 0 & 0 & 0 & 0 & 0 & 0 & 0 & 0 & 0
}
\end{equation}
This point of view is not particularly useful for our purposes because it breaks the $\mathbb{Z}_3$ rotational symmetry
by forcing to distinguish ``incoming'' and ``outgoing'' edges.

Rotate now every tile 120 degrees clockwise: we obtain the tiles $\alpha_\pm$, $\beta_{\pm0}$, $\gamma_-$
(which are exactly the tiles of $\Tm(v)$). This time it is $\gamma_-$ pairs (again, the shaded tiles) which get
a weight of $x$.
Finally, there exists a third possible orientation of rhombi,
obtained from the first set of tiles by 120 degrees counterclockwise rotation 
-- and correspondingly, a third shaded tile which has not been used so far, namely $\gamma_0$. 
We shall use the following graphical notation for these three types of $R$-matrices:
simply depict them with rhombi (with thick edges) and put the spectral parameter $x$ inside. The convention
is that when several rhombi are pasted together, the internal edges are free i.e.\ summed over while the external edges are fixed.

We now have the key proposition, which is the Yang--Baxter equation:
\begin{prop}\label{ybe}
Let three variables $x$, $y$, $z$ satisfy $x+y+z=0$. Then, the following equality holds
\[
\psfrag{=}{$=$}\psfrag{x}{$z$}\psfrag{y}{$x$}\psfrag{z}{$y$}
\myincludegraphics{0.6}{1.15}{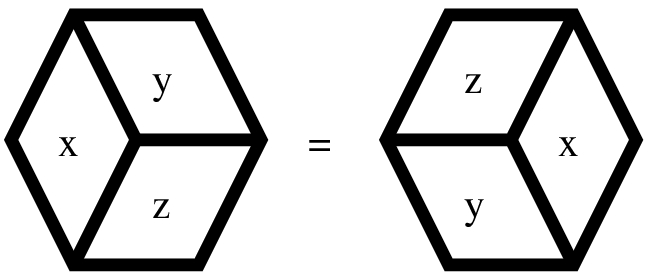}
\]
for any values of the external edges while summing over the matching tiles inside.
\end{prop}
\begin{proof}
The proof is just an explicit computation.
We provide a particularly compact version of it, which was
suggested to the author by A.~Knutson.

Note first that if the equation is true for a given sequence of external
edges, then it is true for any cyclic rotation of it: indeed
a rotation of 120 degrees amounts to cyclic permutation of the variables,
and a rotation of 180 degrees amounts to exchanging l.h.s.\ and r.h.s.
(the tile weights being invariant by 180 degrees rotation).

Next, observe that if no shaded tiles appear in the equality,
then it is trivial -- the different types of rhombi are only distinguished
by the additional shaded tiles which carry a non-trivial weight.
If there are such pieces, then necessarily one of the external sides
of the tiles must be a sequence of $-,+$ (read counterclockwise).
This leaves only three possible sequences of external edges,
up to cyclic rotation: (i)
$-,+,-,-,+,-$, (ii) $-,+,+,-,+,+$ and (iii) $-,+,-,+,-,+$.
The first two sequences are invariant by 180 degrees, which allows
to prove the equality without calculation, since as already mentioned
the 180 degrees rotation of the l.h.s.\ is the r.h.s. 
The last case is the only interesting one: we find the identity
\[\psfrag{+}{$+$}\psfrag{=0}{$=\ 0$}
\myincludegraphics{0.6}{1.15}{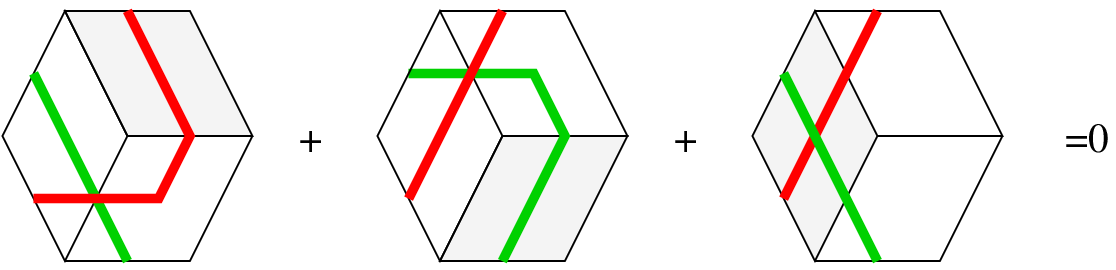}
\]
(or its 180 degrees rotation), that is $x+y+z=0$.
\end{proof}

\subsection{The RTT relations}
We use the notation $\T_\pm^{a,b}(x)$ for the 
transfer matrix that creates one extra row made of the usual tiles $\alpha$, $\beta$ and $\gamma_\pm$,
followed as usual by a $\pm\oh$-step, 
in such a way that sufficiently far to the left the left sides of the rhombi are all $a$ and sufficiently
far to the right the right sides of the rhombi are all $b$. The weights are those of $\R$, 
divided by a factor $\kappa_\pm^{a,b}$ if say the value of the upper edge is equel to $\pm$, which makes sure
that the tiles have weight 1 sufficiently far to the left and right. Explicitly, $\kappa_+^{a,b}=\R_{a,+,a,+}$
and $\kappa_-^{a,b}=\R_{-,b,-,b}$.
In practice we shall only consider here
the cases $\T_\pm^{-,+}$, $\T_+^{-,-}$ and $\T_-^{+,+}$, in which case we recover.
\[
\T_\pm = \T_\pm^{-,+}(0) 
\qquad
\Tp(u)=\T_+^{-,-}(u^{-1})
\qquad
\Tm(u)=\T_-^{+,+}(u^{-1})
\]

We now apply the standard argument of repeated application of the Yang--Baxter equation (``unzipping'') 
to write the RTT relations, which allow to show various commutation relations.

Let us fix two basis elements $\ket{f}$ and $\ket{g}$ in $\G$
and consider matrix elements of the form $\bra{f}\T_+^{a,b}(x)\T_-^{c,d}(y)\ket{g}$. 
The important observation is that to the left of $\min(N_-(f),N_-(g))$, one has series of identical tiles.
Thus no information is lost by truncating 
the state anywhere left of $\min(N_-(f),N_-(g))$, and
the product of weights of the discarded left tiles is 1.
Similarly, to the right of $\max(N_+(f),N_+(g))$, everything is frozen and we only have weights of $1$.

\looseness=-1
Note that we already know this to be the case ``sufficiently far to the left'', but this is not good enough for
our purposes: we need a bound that is uniform w.r.t.\ the intermediate state.
For example if $a=b=-$ and $c=d=+$,
one can check that the infinite sequence of 
$\vcenter{\hbox{\edgerelabelling\myincludegraphics{0.5}{1.15}{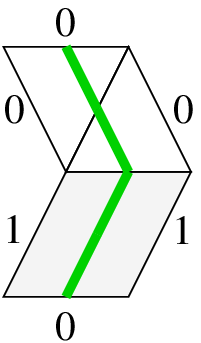}}}$ 
on the left side
and of $\vcenter{\hbox{\edgerelabelling\myincludegraphics{0.5}{1.15}{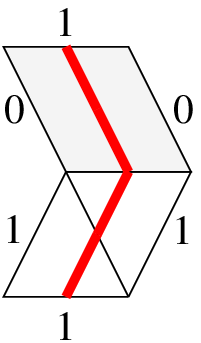}}}$ 
on the right side can only stop when the upper or lower edges change their
value.

\looseness=-1
In the case $a=b=-$ and $c=-$, $d=+$,
there is no change to 
the right of $\max(N_+(f),N_+(g))$, where we find again
sequences of $\vcenter{\hbox{\edgerelabelling\myincludegraphics{0.5}{1.15}{rightbc}}}$. However, to the left of 
$\min(N_-(f),N_-(g))$, we have a new situation where green lines move one full step to the left, that is
sequences of 
$\vcenter{\hbox{\edgerelabelling\myincludegraphics{0.5}{1.15}{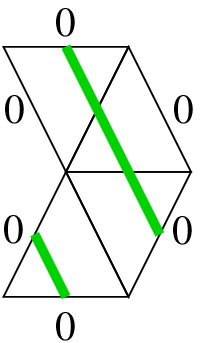}}}$ 
The two other cases of interest to us can be treated similarly.

\newcommand\speclabelling{\def\psfragspec##1{\psfrag{x-t##1}{$\!\scriptstyle\ x$}\psfrag{y-t##1}{$\!\scriptstyle\ y$}}%
\psfragspec{1}\psfragspec{2}\psfragspec{3}\psfragspec{4}\psfragspec{5}\psfragspec{6}\psfrag{-x-y}{$\scriptstyle\ \ z$}}

Thus, if we pick any region containing the interval $[\min(N_-(f),N_-(g)),\max(N_+(f),N_+(g))]$, we can write that
\begin{equation*}
(\kappa_+^{a,b})^{r}(\kappa_-^{c,d})^{g}
\bra{f}\T_+^{a,b}(x)\T_-^{c,d}(y)\ket{g}
=\vcenter{\hbox to 6cm{\hfil $f$\hfil}%
\hbox{\psfrag{a}{$\scriptscriptstyle a$}\psfrag{b}{$\scriptscriptstyle b$}\psfrag{c}{$\scriptscriptstyle c$}\psfrag{d}{$\scriptscriptstyle d$}%
\speclabelling%
\myincludegraphics{0.5}{1.15}{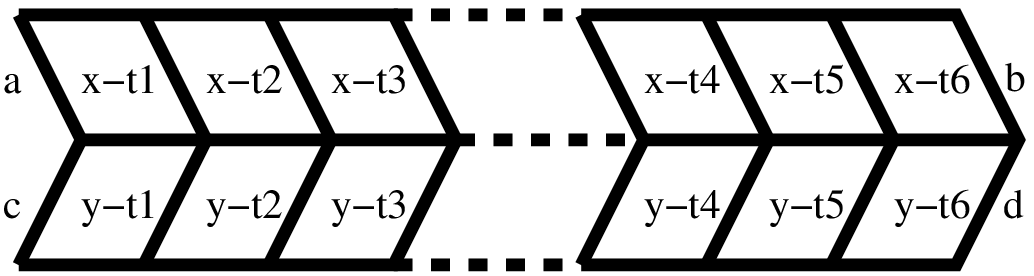}}\hbox to 6cm{\hfil $g$\hfil}}
\end{equation*}
where $g$ (resp. $r$) denotes the number of green (resp.\ red) lines in the finite
domain, which is well-defined if $b=-$ (resp.\ $c=+$) because no green (resp.\ red) line can cross the
boundary with such boundary conditions (and these are the only cases where the factors $\kappa$ are not $1$).

At this stage, we can insert the $R$-matrix at the left, with spectral parameter $z=-x-y$:
\[
\sum_{a,c} \R_{c',a',c,a}(z)
(\kappa_+^{a,b})^g(\kappa_-^{c,d})^{r}
\bra{f}\T_+^{a,b}(x)\T_-^{c,d}(y)\ket{g}
=\vcenter{\hbox to 6.5cm{\hfil $f$\hfil}%
\hbox{\psfrag{a'}{$\scriptscriptstyle a'$}\psfrag{b}{$\scriptscriptstyle b$}\psfrag{c'}{$\scriptscriptstyle c'$}\psfrag{d}{$\scriptscriptstyle d$}%
\speclabelling%
\myincludegraphics{0.5}{1.15}{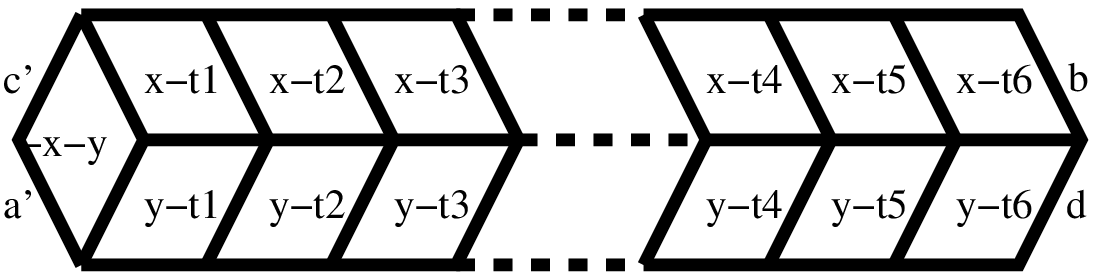}}\hbox to 6.5cm{\hfil $g$\hfil}}
\]
We apply repeatedly proposition \ref{ybe} all the way to 
\begin{multline*}
\sum_{a,c} \R_{c',a',c,a}(z)
(\kappa_+^{a,b})^{r}(\kappa_-^{c,d})^{g}
\bra{f}\T_+^{a,b}(x)\T_-^{c,d}(y)\ket{g}
=\vcenter{\hbox to 6.5cm{\hfil $f$\hfil}%
\hbox{\psfrag{a'}{$\scriptscriptstyle a'$}\psfrag{b}{$\scriptscriptstyle b$}\psfrag{c'}{$\scriptscriptstyle c'$}\psfrag{d}{$\scriptscriptstyle d$}%
\speclabelling%
\myincludegraphics{0.5}{1.15}{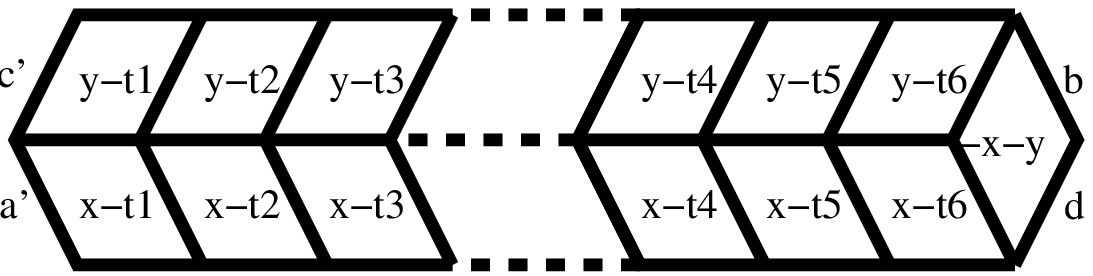}}\hbox to 6.5cm{\hfil $g$\hfil}}
\\
=\sum_{b',d'}\R_{d',b',d,b}(-x-y) 
(\kappa_+^{a',b'})^{r}(\kappa_-^{c',d'})^{g}
\bra{f}
\T_-^{c',d'}(y)\T_+^{a',b'}(x)
\ket{g}
\end{multline*}

By varying $a',b,c',d$ one gets in principle 81 identities relating transfer matrices with various boundary
conditions. We are only interested in the four cases $a'=-$, $b\in\{+,-\}$, $c'\in\{+,-\}$, 
$d=+$ which lead to the following

\begin{theorem}
We have the commutation relations
\begin{equation*}
[\T_+^{-,-}(x),\T_-^{-,+}(y)]
=
[\T_+^{-,-}(x),\T_-^{+,+}(y)]
=
[\T_+^{-,+}(x),\T_-^{-,+}(y)]
=
[\T_+^{-,+}(x),\T_-^{+,+}(y)]
=0
\end{equation*}
\end{theorem}
This theorem implies that $[\Tp(u),\Tm(v)]=0$, as well as $[\tilde\T_{\epsilon}(u),\T_{\epsilon'}]=0$, $\epsilon,\epsilon'\in\{-,+\}$.

\subsection{Other cases}\label{otherR}
\begin{figure}
\psfrag{x}{$x$}\psfrag{y}{$y$}\psfrag{z}{$z$}
\myincludegraphics{0.6}{1.15}{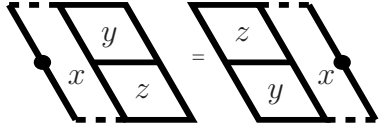}%
\caption{Graphical representation of the Yang--Baxter equation with 
one matrix $\R^\star(x)$ and two matrices $\R(y)$, $\R(z)$. Here $x=z-y$.}
\label{figotherybe}
\end{figure}
In order to complete the proof of theorem \ref{mainthm}, we must show
the commutation $[\Tpm(u),\Tpm(v)]=0$ (same sign). This requires another kind of $R$-matrix, as the geometry
suggests, cf Fig.~\ref{figotherybe}. We do not provide the details since these commutation relations are not needed anywhere.
Let us simply give the expression of the new $R$-matrix:
\[
\R^{\star}(x)=\kbordermatrix{
&-,-&-,+&-,0&+,-&+,+&+,0&0,-&0,+&0,0\\
-,-& 1 & 0 & 0 & 0 & 0 & 0 & 0 & 0 & 0 \\
-,+& 0 & x & 0 & 1 & 0 & 0 & 0 & 0 & 0 \\
-,0& 0 & 0 & x & 0 & 0 & 0 & 1 & 0 & 0 \\
+,-& 0 & 1 & 0 & 0 & 0 & 0 & 0 & 0 & 0 \\
+,+& 0 & 0 & 0 & 0 & 1 & 0 & 0 & 0 & 0 \\
+,0& 0 & 0 & 0 & 0 & 0 & 0 & 0 & 1 & 0 \\
0,-& 0 & 0 & 1 & 0 & 0 & 0 & 0 & 0 & 0 \\
0,+& 0 & 0 & 0 & 0 & 0 & 1 & 0 & x & 0 \\
0,0& 0 & 0 & 0 & 0 & 0 & 0 & 0 & 0 & 1
}
\]
The rest of the proof is identical (unzipping argument) and we obtain the following
\begin{theorem}
We have the commutation relations
\begin{equation*}
[\T_+^{-,-}(x),\T_+^{-,-}(y)]
=
[\T_-^{+,+}(x),\T_-^{+,+}(y)]
=
[\T_+^{-,+}(x),\T_+^{-,+}(y)]
=
[\T_-^{-,+}(x),\T_-^{-,+}(y)]
=0
\end{equation*}
\end{theorem}

Note that this theorem does {\em not}\/ say that $\T_+^{-,-}$ and $\T_+^{-,+}$ commute.

\section{Littlewood--Richardson coefficients from the coproduct}\label{nonequiv}
\subsection{The coproduct formula}
\begin{theorem}\label{finalthm}
The coefficients $c^{\lambda}_{\mu,\nu}$ in the corollary of
lemma \ref{freeactionb} are Littlewood--Richardson coefficients.
\end{theorem}
\begin{proof}
Start from the supersymmetric Schur function
(lemma \ref{basicrel})
\[
s_\lambda(u_1,\ldots,u_m/v_1,\ldots,v_n)
=\bra{\varnothing} \prod_{i=1}^m \Tf(u_i)\prod_{i=1}^n \Tff(v_i) \ket{\lambda}
\]

Considering $\ket{\varnothing}$ and $\ket{\lambda}$ as states in $\G$,
we immediately find by applying lemma \ref{crossaction}:
\[
s_\lambda(u_1,\ldots,u_m/v_1,\ldots,v_n)
=\bra{\varnothing} \prod_{i=1}^m \Tp(u_i)\prod_{i=1}^n \Tm(v_i) \ket{\lambda}
\]

Given a non-negative integer $k$, consider the evolution {\em backwards in time} (i.e.\ going downwards instead of upwards) starting from
$\bra{\varnothing}\concat_k\bra{\varnothing}$. It is elementary to check that after $2k$ steps, one gets $\bra{\varnothing}$,
or in other words
$\bra{\varnothing}=\bra{\varnothing}\concat_k\bra{\varnothing}\T^{2k}$. This results in
\[
s_\lambda(u_1,\ldots,u_m/v_1,\ldots,v_n)
=\bra{\varnothing}\concat_k\bra{\varnothing} 
\T^{2k} \prod_{i=1}^m \Tp(u_i)\prod_{i=1}^n \Tm(v_i) \ket{\lambda}
\]

We now use the commutation of the transfer matrices (theorem \ref{mainthm}) to rewrite it
\[
s_\lambda(u_1,\ldots,u_m/v_1,\ldots,v_n)
=
\bra{\varnothing}\concat_k\bra{\varnothing} 
\prod_{i=1}^m \Tp(u_i)\prod_{i=1}^n \Tm(v_i) \T^{2k}\ket{\lambda}
\]
Assume $k>\max(N_{+}(\lambda),-N_{-}(\lambda))$, and apply the corollary of lemma \ref{freeactionb}:
\[
s_\lambda(u_1,\ldots,u_m/v_1,\ldots,v_n)
=
\sum_{\mu,\nu} c^{\lambda}_{\mu,\nu} 
\bra{\varnothing}\concat_k\bra{\varnothing} 
\prod_{i=1}^m \Tp(u_i)\prod_{i=1}^n \Tm(v_i) 
\ket{\mu}\concat_k\ket{\nu}
\]
We now wish to apply lemma \ref{freeactionc}. 
$\ket{\mu}\concat_k\ket{\nu}\in \Gf$, but as already mentioned
$\Tp(u)$ may push red particles to the left and $\Tm(u)$ may push green particles to the right.
Let us thus choose $k>\max(N_{+}(\lambda),-N_{-}(\lambda))+n/2$; according to lemma \ref{motionb}, this ensures that
lemma \ref{freeactionc} can be applied repeatedly.
Separating into two bra-kets, we get:
\[
s_\lambda(u_1,\ldots,u_m/v_1,\ldots,v_n)
=
\sum_{\mu,\nu} c^{\lambda}_{\mu,\nu} 
\bra{\varnothing}
\prod_{i=1}^m \Tf(u_i)
\ket{\nu}
\bra{\varnothing}
\prod_{i=1}^n \Tff(v_i)
\ket{\mu}
\]
Finally, apply the basic free fermionic identity (lemma \ref{basicrel}): we find
\[
s_\lambda(u_1,\ldots,u_m/v_1,\ldots,v_n)
=
\sum_{\mu,\nu} c^{\lambda}_{\mu,\nu} \,
s_{\nu}(u_1,\ldots,u_m)
s_{\mu^T}(v_1,\ldots,v_n)
\]
Remembering that the l.h.s.\ is the supersymmetric Schur function associated to $\lambda$, we conclude that
this equality defines uniquely
the $c^\lambda_{\mu,\nu}$ as Littlewood--Richardson coefficients.
\end{proof}

\subsection{Back to the triangle}\label{backtri}
\begin{figure}
\psfrag{N-}{$\scriptstyle N_-(\lambda)$}\psfrag{N+}{$\scriptstyle N_+(\lambda)$}%
\myincludegraphics{0.5}{1.15}{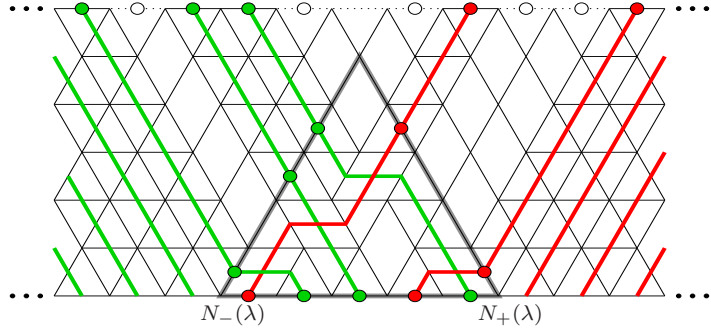}
\caption{An example of configuration and the triangle in which all the crossings take place.}\label{figtriangle}
\end{figure}
All that has been described so far uses an infinite-dimensional Fock space of configurations. 
This helps in formulating the results more elegantly.
However,
for any given Young diagram $\lambda$, all the evolution of the state $\ket{\lambda}$ with respect to $\T$ takes place in a finite
part of the plane; more precisely, a triangle, which leads us back to the more standard formulation, as in \cite{KT},
in terms of puzzles.

First, let us define how to read Young diagrams from the boundaries of a finite domain (this is
a finitized version of the transformation of section \ref{fockfree}).
Let $n$ and $k$ be two non-negative integers such that $k\le n$. 
Consider a sequence of $n$ successive edges which can be either $-$ or $+$, such that
there are $k$ ``$-$'' and $n-k$ ``$+$''. Fix an orientation of the
edge: then reading the values of the edges following the orientation, one obtains a binary string of length $n$.
To it one associates bijectively 
a Young diagram contained inside the Young diagram $k\times (n-k)$, that is the rectangle
of height $k$ and width $n-k$, as follows: each $-$ corresponds to a step up, and each $+$ to a step to the right,
starting from the lower left corner of the rectangle. For example,
\[
+-++-+\ \longmapsto\ 
\vcenter{\hbox{\setlength{\unitlength}{\cellsize}%
\begin{picture}(4,2)%
\put(0,0){\line(1,0){4}}%
\put(0,0){\line(0,1){2}}%
\put(0,2){\line(1,0){4}}%
\put(4,0){\line(0,1){2}}%
\linethickness{2pt}%
\put(0,0){\line(1,0){1}}%
\put(1,0){\line(0,1){1}}%
\put(1,1){\line(1,0){1}}%
\put(2,1){\line(1,0){1}}%
\put(3,1){\line(0,1){1}}%
\put(3,2){\line(1,0){1}}%
\end{picture}}}
\ \equiv\ \tableau{&&\\\\}
\]

Note that depending on the direction of
the boundary, a $-$ will be drawn as a green particle or an empty spot, and a $+$ will be drawn as a red particle or 
an empty spot. 

Let us now fix an initial state given by a Young diagram $\lambda$ and consider the possible evolution of the
system.
The situation is illustrated on Fig.~\ref{figtriangle}. The smallest possible triangle 
in which all crossings between green and red particles take place 
has as its lower side the interval
between $N_-(\lambda)$ and $N_+(\lambda)$. So we can set $n=N_+(\lambda)-N_-(\lambda)$, in which case
$k$ is the height of $\lambda$ (number of non-zero parts). 
Of course one could make the triangle bigger,
which illustrates the stability property of puzzles.
Now the configuration outside the triangle is uniquely fixed by the locations of the lines at the two upper
sides of the triangle: green (resp.\ red) particles move uniformly to the left
(resp.\ right). In other words, no information is lost by restricting to the triangle and
conversely, any configuration inside the triangle can be extended to the outside in a unique way.
The ``asymptotic'' states $\mu$ and $\nu$ which describe the sequences of green particles
and empty spots to the left and of red particles and empty spots to the right can also be read off the two
upper sides of the triangle in the way described in the previous paragraph, the binary strings being read from
left to right. 
Combining these observations, we conclude that $c^\lambda_{\mu,\nu}$ also counts
the number of fillings of the triangle with fixed boundaries,
where $\lambda$ is encoded by the bottom edge,
$\mu$ by the left edge and
$\nu$ by the right edge, always reading from left to right. These fillings are called {\em puzzles}.
For example, on Fig.~\ref{figtriangle}, one reads
\[
\medboxes
\lambda=\tableau{&\\ \\ \\}\,,\ 
\mu=\nu=\tableau{\\ \\}\,.
\]


\section{Equivariance or the introduction of inhomogeneities}\label{equiv}
It is natural to try to generalize the construction of section \ref{nonequiv} by putting arbitrary
horizontal and vertical spectral parameters (inhomogeneities) at each site. 
As already observed in \cite{artic32,artic33,artic34} in a different but similar setting,
this amounts geometrically  to going from ordinary cohomology to equivariant cohomology -- in the present
case, of the Gra\ss mannian.

There are however some complications.
The first one is that the infinite system viewpoint used so far is not particularly convenient to handle the functions that
appear in the inhomogeneous situation, that is factorial (or double) Schur functions.
We shall therefore work in a finite domain instead.

The second one is that the coproduct formula does not obviously generalize to the inhomogeneous case. In fact,
a coproduct formula for double Schur functions has only recently been discovered \cite{Molev-coproduct} and the description 
in terms of objects analogous to puzzles is unknown. We shall prove instead product formulae.
Note that even specialized to the non-equivariant case, the formulae of this section will thus be distinct from those of section
\ref{nonequiv}.

The basis setup is to consider that there are (oriented) lines of spectral parameters propagating on the 
{\em dual lattice} of the lattice of tiles of $R$-matrices, in such a way that the spectral parameter attached
to the tile is the difference of the two spectral parameters crossing it:
\[
\psfrag{y}{$y$}\psfrag{z}{$z$}\psfrag{z-y}{$y-z$}\psfrag{=}{$=$}
\myincludegraphics{0.8}{1.15}{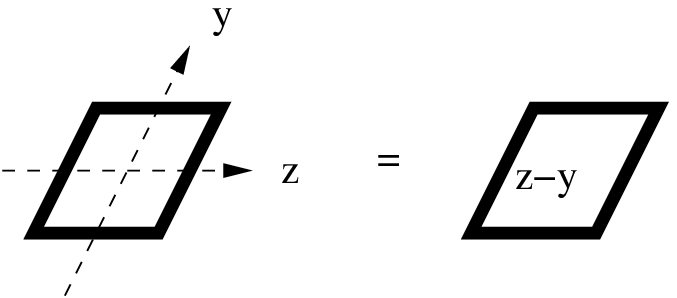}%
\]

In all that follow we fix integers $n$ and $k$, $k\le n$, and use the same correspondence between
binaray strings and Young diagrams inside the rectangle $k\times (n-k)$ described in section \ref{backtri}.

We now define the two basic ``building blocks'' with which we shall produce non-trivial identities.

\subsection{Factorial Schur functions}\label{facschur}
\begin{figure}
\psfrag{l}{$\lambda$}%
\psfrag{k}[Bc]{$\overbrace{\hbox to 2.7cm{\hfill}}^k$}%
\psfrag{n}[cc][][1][63]{$\underbrace{\hbox to 3.5cm{\hfill}}$}\psfrag*{n}[lc]{\quad$n$}%
\psfrag{y1}{$\scriptstyle y_1$}\psfrag{y2}{$\scriptstyle y_2$}\psfrag{y3}{$\scriptstyle y_3$}\psfrag{y4}{$\scriptstyle y_4$}\psfrag{y5}{$\scriptstyle y_5$}%
\psfrag{x1}{$\scriptstyle x_1$}\psfrag{x2}{$\scriptstyle x_2$}\psfrag{x3}{$\scriptstyle x_3$}\psfrag{x4}{$\scriptstyle x_4$}%
\myincludegraphics{0.5}{1.15}{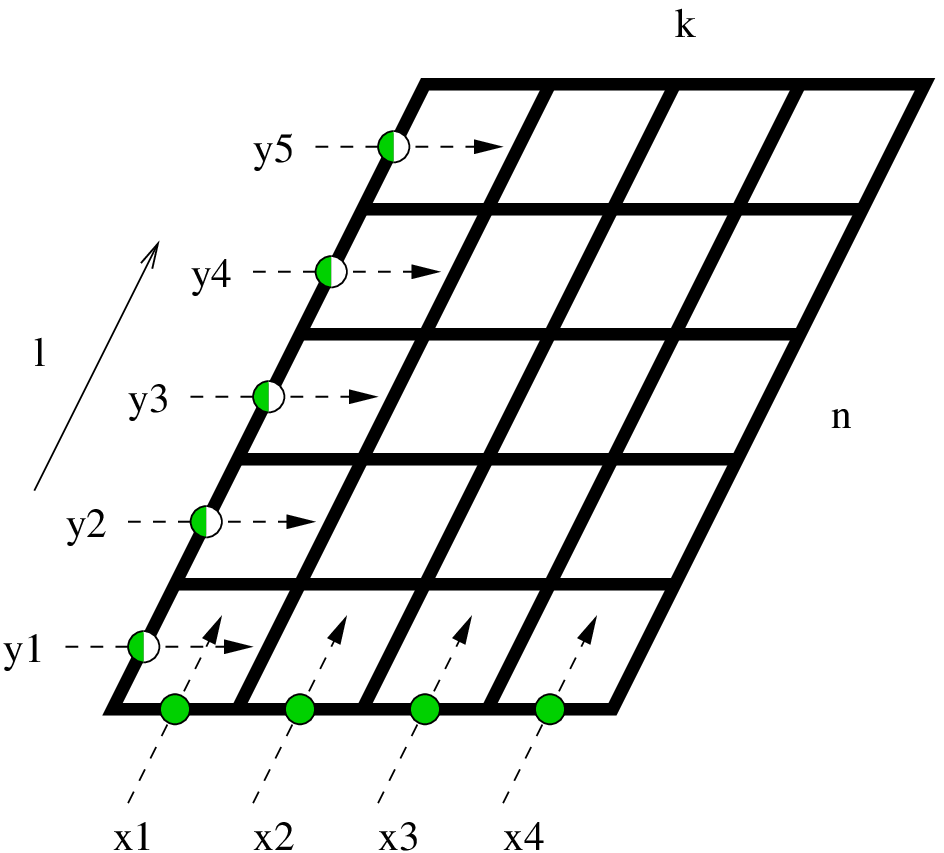}%
\quad%
\psfrag{l}{$\lambda$}%
\myincludegraphics{0.5}{1.15}{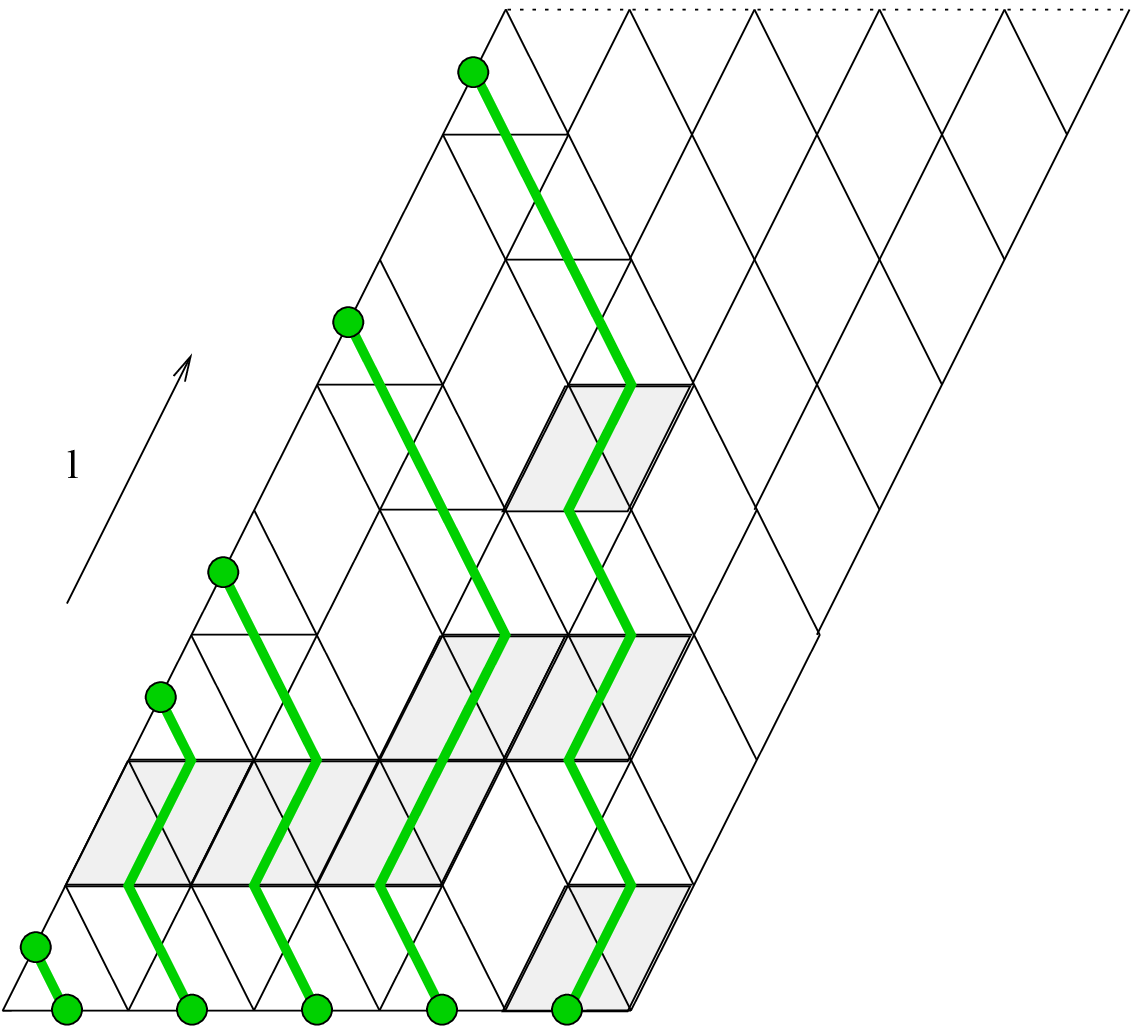}%
\par\vskip1cm
\hskip9cm\tableau{1&2&3\\3&3\\4\\5\\}
\vskip-2.5cm
\includegraphics{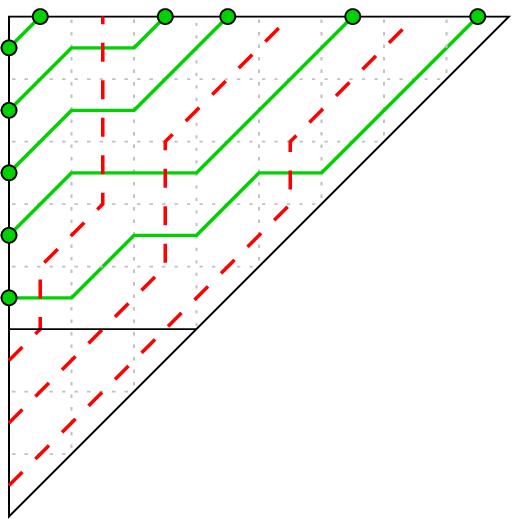}%
\caption{Our definition of factorial Schur functions, and an example in various graphical representations. The numbering of the $x_i$ is reversed
in the tableau representation.}\label{figdoubleschur}
\end{figure}
What we really want to define is the double
Schubert polynomial of a Gra\ss mannian permutation. It will however
coincide with the notion of factorial Schur function.

Given a Young diagram $\lambda$ inside the rectangle $k\times(n-k)$,
let us define the factorial Schur function $s_\lambda(x_1,\ldots,x_k|y_1,\ldots,y_n)$ 
graphically as on Fig.~\ref{figdoubleschur}, that is as a tiling
of a $n\times k$ rhombus with prescribed boundaries: $\lambda$ is encoded
into the edges of the left side, read from bottom to top, the bottom side
is full of green particles and the other two sides are empty (in all the figures of this section,
we omit entirely drawing empty spots, and use half-colored circles to indicate that the spot can be either
empty or occupied by a particle of the given color, depending on the Young diagram it encodes).
The spectral parameters are the $x_i$ and the $y_i$, from left to right and bottom to top.

An important property is the following:
\begin{prop}\label{doubleschursym}
$s_\lambda(x_1,\ldots,x_k|y_1,\ldots,y_n)$ is a symmetric function of the $x_i$, $i=1,\ldots,k$.
\end{prop}
\begin{proof}
Fix $i\in \{1,\ldots,n-1\}$ and
apply the matrix $\R^\star(x_{i+1}-x_i)$ 
from section \ref{otherR} to the bottom edges
$i$ and $i+1$. Noting that $\R^\star_{i,j,-,-}=\delta_{i,-}\delta_{j,-}$
and $\R^\star_{i,j,0,0}=\delta_{i,0}\delta_{j,0}$,
we can use the usual unzipping argument (repeated application of
the Yang--Baxter relation represented on Fig.~\ref{figotherybe}) to move
the matrix $\R^\star(x_{i+1}-x_i)$ all the way to the top and then remove it.
The result is the same picture as we started from, but with $x_i$ and
$x_{i+1}$ exchanged.
\end{proof}

On Fig.~\ref{figdoubleschur}, an example of configuration is provided, 
as well as two alternative representations of it.
The first one is the {\em pipe dream}\/ of the Gra\ss mannian permutation
associated to $\lambda$ \cite{FK-Schubert,KM-Schubert}. 
Our picture
can be deformed into the pipe dream picture by rotating 60 degrees clockwise
and then distorting it slightly to make the 60 degrees angle a right angle.
This way, the green lines
are exactly the trajectories of the lines above the descent of the
permutation. The other lines can be recovered unambiguously. The weights
are now given to the crossings (between a red line and a green line) and
take the form $x_i-y_j$, where $i$ is the row number and $j$ the column number,
both counted starting from the corner.

The final representation is a semi-standard Young tableau of $\lambda$,
with the alphabet $\{1,\ldots,k\}$.
Starting from the pipe dream picture, each column of the tableau
corresponds to a red line
counted from left to right, and each number in the boxes of the column
indicates the row at which the red line crosses a green line. An important
difference is that the rows are numbered from bottom to top;
but since the factorial Schur function is a symmetric function of the
$x_i$, this is irrelevant, and we may as well assign to the
tableau a weight equal to the product over boxes of $x_T-y_{T+c}$, 
where $T$ is the number in the box 
(i.e.\ the row number in the pipe dream picture)
and $c$ is the content of the box (column minus row of the box),
in such a way that $T+c$ is precisely the column number in the pipe dream
picture.

\subsection{MS-puzzles and equivariant puzzles}\label{puzzles}
\begin{figure}
\psfrag{theta}{$\lambda$}\psfrag{mu}{$\mu$}\psfrag{nu}{$\nu$}%
\psfrag{y1}{$\scriptstyle y_1$}\psfrag{y2}{$\scriptstyle y_2$}\psfrag{y3}{$\scriptstyle y_3$}\psfrag{y4}{$\scriptstyle y_4$}\psfrag{y5}{$\scriptstyle y_5$}%
\psfrag{z1}{$\scriptstyle z_1$}\psfrag{z2}{$\scriptstyle z_2$}\psfrag{z3}{$\scriptstyle z_3$}\psfrag{z4}{$\scriptstyle z_4$}\psfrag{z5}{$\scriptstyle z_5$}%
\psfrag{n1}[cc][][1][63]{$\ \overbrace{\hbox to 2.6cm{\hfill}}$}\psfrag*{n1}[rb]{$n\,$}%
\psfrag{n2}[cc][][1][117]{$\underbrace{\hbox to 2.6cm{\hfill}}$}\psfrag*{n2}[lc]{\quad$n$}%
\myincludegraphics{0.5}{1.15}{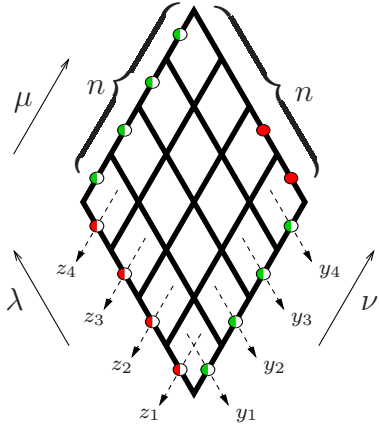}%
\caption{Definition of an MS-puzzle ($k=2$, $n=4$).}
\label{figdefms}
\end{figure}
We first define an MS-puzzle (this terminology is borrowed from \cite{KT}).
It is simply given by Fig.~\ref{figdefms}. Three sides encode Young diagrams, whereas the fourth side, the upper right one, is simply a sequence of $+^{n-k}-^k$, all read from bottom to top.
We denote this object by $e^\nu_{\lambda,\mu}(y_1,\ldots,y_n;z_1,\ldots,z_n)$.

We shall need the following series of lemmas:
\begin{lemma}\label{prefrozen}
In the top half of a MS-puzzle, green lines always go in straight lines
(up and to the left).
\end{lemma}
\begin{proof}
By inspection.
Since the red lines are stacked at their rightmost positions, 
there is no available free spot for a green particle to cross the red particles
by moving straight to the right.
\end{proof}

\begin{lemma}\label{frozena}
Assume that $z_{n+1-i}=y_i$, $i=1,\ldots,n$.
Then the top half of the MS-puzzle is ``frozen'' i.e.\ it has a unique configuration (which has
weight 1);
the edges on the horizontal diagonal of the MS-puzzle reproduce, read from left to right, the Young diagram $\mu$.
\end{lemma}
\begin{proof}
If $z_{n+1-i}=y_i$ for all $i$, then the rhombi sitting on the horizontal diagonal of the MS-puzzle have a zero spectral parameter. This implies that the
shaded tiles $\gamma_0$ is forbidden, or equivalently that the edges crossing horizontally these rhombi cannot be $\tilde 0$ (red and green particles on top of each other). Since there are $k$ green and $n-k$ red lines incoming, there
cannot be an empty spot either (edge $0$). So these edges can only
be $+$ or $-$; according to lemma \ref{prefrozen}, the green particles
on it are at the same locations as on the upper left edge, and therefore
the red lines are also fixed (they must move one step to the left each time they cross a green line) and occupy the complementary set on the diagonal. 
In other words, the diagram $\mu$ is reproduced on the diagonal.
No shaded tiles are used in the top half, so its weight is one.
\end{proof}

\begin{figure}
\psfrag{theta}{$\lambda$}\psfrag{mu}{$\mu$}\psfrag{nu}{$\nu$}%
\psfrag{mubar}{$\bar\mu$}\psfrag{nubar}{$\bar\nu$}%
\myincludegraphics{0.5}{1.15}{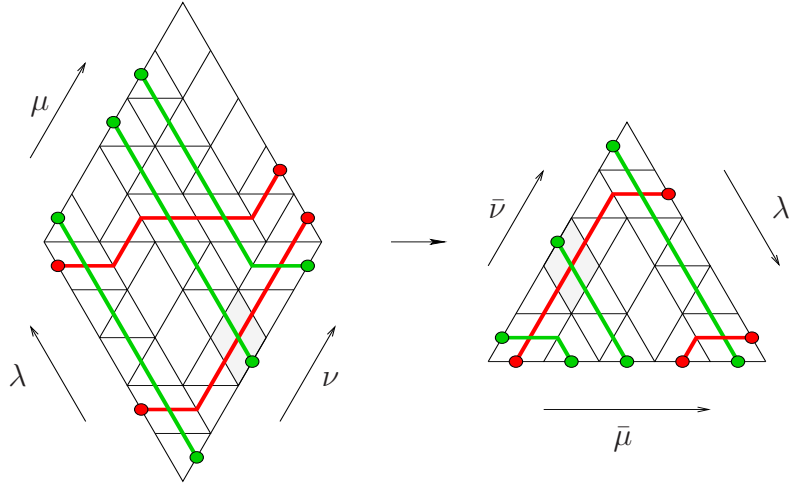}%
\caption{From an MS-puzzle to an equivariant puzzle.}
\label{figequivpuzzle}
\end{figure}

With the hypothesis of the last lemma, the top half of the MS-puzzle
can be removed since it is fixed and has a weight of 1. The result,
after 180 degree rotation, is called an {\em equivariant puzzle}, cf
Fig.~\ref{figequivpuzzle}.

If we call $c^\nu_{\mu,\lambda}$ the weight of an equivariant puzzle, that is
the tiling of a triangle with sides $\nu$ (bottom), $\mu$ (left), $\lambda$ (right), all read
from left to right, with the same tiles and spectral parameters
as an upside-down MS-puzzle with $z_{n+1-i}=y_i$, we have
\[
e^\nu_{\lambda,\mu}(z_n,\ldots,z_1;z_1,\ldots,z_n)=
c^{\bar\mu}_{\bar\nu,\lambda}(z_1,\ldots,z_n)
\]
or equivalently $c^\nu_{\mu,\lambda}(z_1,\ldots,z_n)=e^{\bar\mu}_{\lambda,\bar\nu}(z_n,\ldots,z_1;z_1,\ldots,z_n)$,
where $\bar\lambda$ is the complement of the Young diagram $\lambda$ inside the rectangle
$k\times(n-k)$ after a 180 degree rotation, which corresponds to reading the binary string of $\lambda$ from right to left.

\begin{lemma}\label{frozenb}
There are no equivariant puzzles contributing
to $c^\nu_{\varnothing,\mu}$ or $c^\nu_{\mu,\varnothing}$ if
$\mu\ne\nu$, and one (which has no shaded tiles) if $\mu=\nu$, so that:
\[
c^\nu_{\varnothing,\mu}(z_1,\ldots,z_n)=c^\nu_{\mu,\varnothing}(z_1,\ldots,z_n)=\delta^\nu_\mu
\]
\end{lemma}
\begin{proof}
Considering the equivariant puzzle as the bottom piece of an MS-puzzle,
apply lemma \ref{frozena} to either the mirror image w.r.t.\ the horizontal axis of the MS-puzzle with colors interchanged, 
or its 180 degree rotation.
\end{proof}

Finally, note that if all spectral parameters are equal to zero, 
we find by definition
$c^{\nu}_{\lambda,\mu}(0,\ldots,0)=c^{\nu}_{\lambda,\mu}$; which implies
$e^{\nu}_{\lambda,\mu}(0,\ldots,0;0,\ldots,0)=c^{\bar\mu}_{\bar\nu,\lambda}$.
Rotating a (non-equivariant) puzzle 60 degrees counterclockwise results in another
puzzle, so that 
$c^{\bar\mu}_{\bar\nu,\lambda}=c^{\nu}_{\lambda,\mu}$.
To summarize,
\[
c^{\nu}_{\lambda,\mu}(0,\ldots,0)=
e^{\nu}_{\lambda,\mu}(0,\ldots,0;0,\ldots,0)=c^{\nu}_{\lambda,\mu}
\]

\subsection{The Molev--Sagan problem}
The Molev--Sagan (MS) problem consists in expanding the product of two factorial Schur functions with the same {\em first}
set of variables as a sum of factorial Schur functions. It was first solved in \cite{MS-doubleschur} in terms of barred tableaux
and then reformulated in \cite{KT} in terms of MS-puzzles. We now rederive it in our framework.

\begin{figure}
\psfrag{k}[Bc]{$\overbrace{\hbox to 1.6cm{\hfill}}^k$}%
\psfrag{n}[cc][][1][63]{$\underbrace{\hbox to 2.6cm{\hfill}}$}\psfrag*{n}[lc]{\quad$n$}%
\psfrag{n2}[cc][][1][117]{$\underbrace{\hbox to 2.6cm{\hfill}}$}\psfrag*{n2}[lc]{\quad$n$}%
\psfrag{=}[][][1.5]{$=$}\psfrag{theta}{$\lambda$}\psfrag{mu}{$\mu$}%
\psfrag{y1}{$\scriptstyle y_1$}\psfrag{y2}{$\scriptstyle y_2$}\psfrag{y3}{$\scriptstyle y_3$}\psfrag{y4}{$\scriptstyle y_4$}\psfrag{y5}{$\scriptstyle y_5$}%
\psfrag{z1}{$\scriptstyle z_1$}\psfrag{z2}{$\scriptstyle z_2$}\psfrag{z3}{$\scriptstyle z_3$}\psfrag{z4}{$\scriptstyle z_4$}\psfrag{z5}{$\scriptstyle z_5$}%
\psfrag{x1}{$\scriptstyle x_1$}\psfrag{x2}{$\scriptstyle x_2$}\psfrag{x3}{$\scriptstyle x_3$}%
\myincludegraphics{0.5}{1.15}{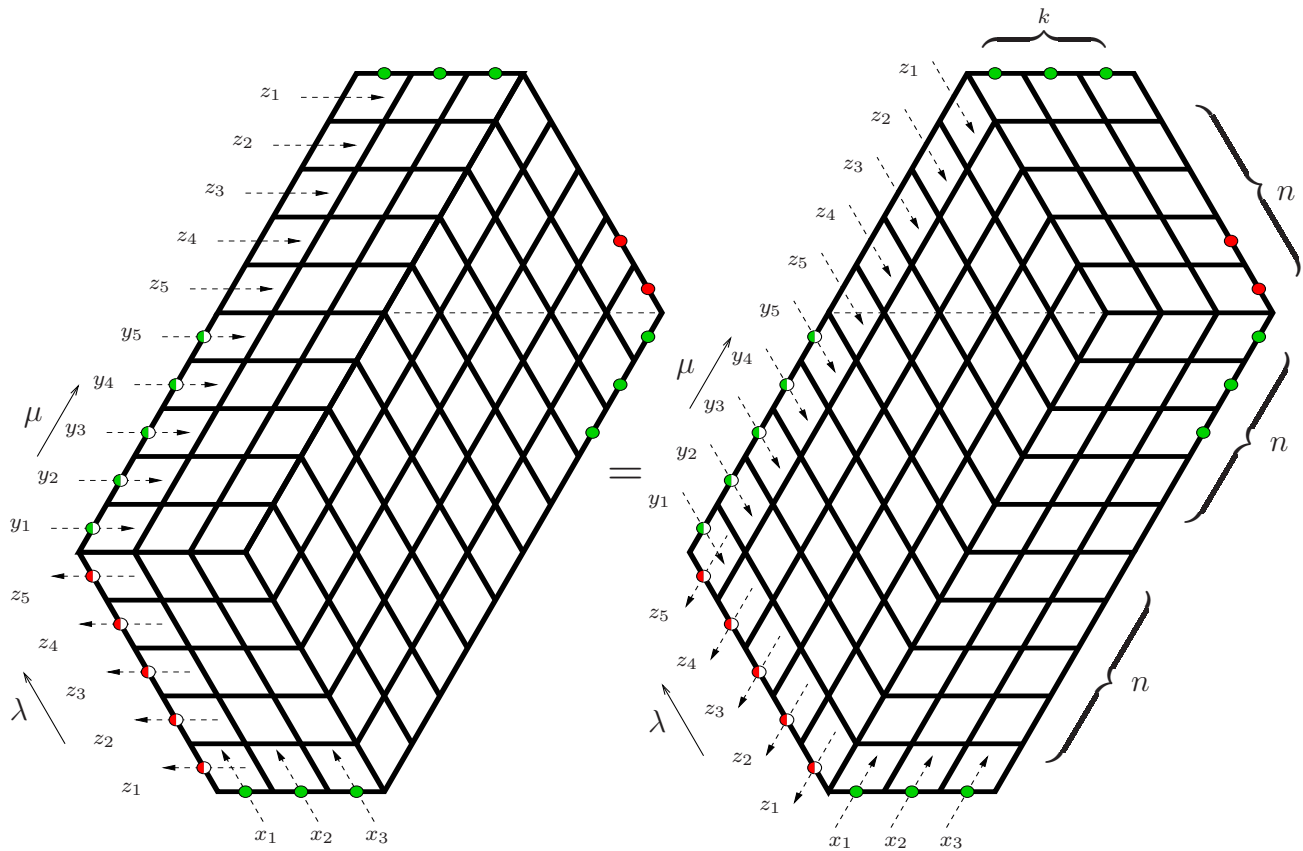}%
\caption{Graphical solution of the MS problem ($n=5$, $k=3$).}
\label{fighexams}
\end{figure}

Let us consider the formal equality of Fig.~\ref{fighexams}. On each side of the equality, we have configurations
where all the external edges are fixed. The order of the parameters $y_i$ and $z_i$ is important; note in particular 
the dashed line which corresponds to the difference of spectral parameters vanishing.

On the upper left side, we have a Young diagram $\mu$ encoded by a binary
string of green particles and empty spots in the region of parameters $y_i$,
and emptiness above in the region of the $z_i$.
On the lower left side, we have a Young diagram $\lambda$ encoded by a binary
string of green particles and empty spots.
Both diagrams are read from bottom to top.
The right sides have $k$ green particles at their highest possible location
and $n-k$ red particles at their lowest possible location.
The top and bottom sides are full of green particles.
Furthermore, we assume that the sum of widths $\mu_1+\lambda_1$ of the Young diagrams $\mu$ and $\lambda$
does not exceed $n-k$. This can always be achieved by choosing $n$ sufficiently large.
The justification of this assumption will become clear below.

In order to go from one side to the other side of the figure, one simply applies repeatedly proposition \ref{ybe} (Yang--Baxter)
starting from the center and ``piling up boxes''. Since our spectral parameters are differences of 
the parameters of crossing lines, the sum of spectral parameters around a hexagon is indeed zero.

\begin{figure}
\psfrag{theta2}{$\smallboxes\mskip-40mu\lambda=\tableau{\\ \\}$}\psfrag{mu2}{$\smallboxes\mskip-45mu\mu=\tableau{\\ \\}$}
\psfrag{theta}{$\lambda$}\psfrag{mu}{$\mu$}\psfrag{nu}{$\nu$}%
\psfrag{A}{A}\psfrag{B}{B}\psfrag{C}{C}\psfrag{D}{\color{blue} D}\psfrag{E}{\color{blue} E}\psfrag{F}{\color{blue} F}\psfrag{G}{G}\psfrag{H}{H}\psfrag{I}{I}\psfrag{J}{\color{blue} J}%
\myincludegraphics{0.5}{1.15}{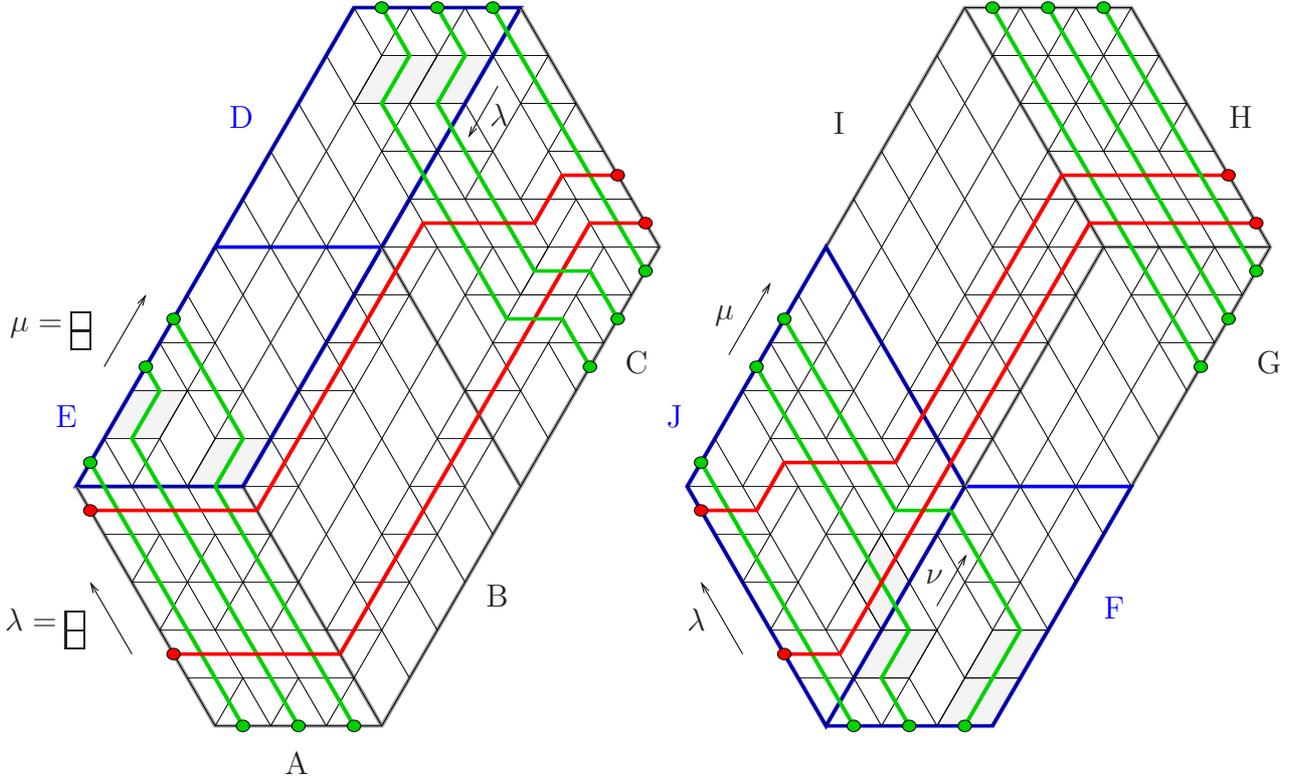}%
\caption{An example of tilings contributing to each side of the equality of Fig.~\ref{fighexams}.}
\label{fighexamsex}
\end{figure}

We shall now examine what the consequences of this equality are. It is simpler to work on an example, as on
Fig.~\ref{fighexamsex}. For ease of discussion, we have labelled the various regions inside the hexagon
and marked in blue the ``interesting'' ones, that is those in which there is some freedom left -- the other regions
are entirely ``frozen'' once the boundary conditions are fixed.

Let us start with the left hand side. In region A, green particles starting from the bottom can only go straight to the left, so that
red particles must go horizontally. In region B, red particles move freely to the right. In region C, 
we note that the spectral parameters are of the form of the hypothesis of lemma \ref{frozena},
so that we can apply lemmas \ref{frozena} and \ref{frozenb} to
conclude that the whole region is frozen.
The result is that the upper left side of region C reproduces the
Young diagram $\lambda$ read from top to bottom. We now recognize region D as the 180 degree rotated picture of a
factorial Schur function (cf section \ref{facschur}). Since all tiles and weights are 180 degree rotation invariant,
we find that this region contributes $s_\lambda(x_1,\ldots,x_k|z_1,\ldots,z_n)$, paying attention to the order
of the parameters $z_i$ (we have however reordered the $x_i$ because of proposition \ref{doubleschursym}).
Region E is the picture of a factorial Schur function with the standard orientation,
so it contributes $s_\mu(x_1,\ldots,x_k|y_1,\ldots,y_n)$. 
Overall, the left hand side equals $s_\lambda(x_1,\ldots,x_k|z_1,\ldots,z_n)s_\mu(x_1,\ldots,x_k|y_1,\ldots,y_n)$.

Now let us compute the right hand side. In regions G, H and I, it is easy to check
that all particles can only move straight in the way indicated on the figure.
Region F needs to be dealt with carefully. We want to make sure that the green lines starting from the bottom
exit region F through the left side. This requires to look ahead into the region J.
Let us consider the rightmost green particle. If we number the upper left side of region J from bottom to top,
then its endpoint is precisely $k+\mu_1$ (the length of the first row of $\mu$).
This means that the location (counted from bottom to top) of the rightmost green particle when it crosses
the diagonal line starting at the bottom of the junction of J and F cannot exceed $k+\mu_j$ plus the number
of steps to the left it can make in region J. We claim that this number is $\lambda_1$: indeed
among the $n-k$ red particles starting from the lower left side of J, 
only $\lambda_1$ are allowed to cross without making a step to the right.
The last $n-k-\lambda_1$, which are the topmost ones, have to make a step to the right each time they
cross a green particle in order to reach their final destination to the right of the upper right side of F.
The result is that this location is at most $k+\mu_1+\lambda_1$, which by assumption is less or equal to $n$.
Thus the rightmost green particle, and therefore all others, exit through the left side. We now recognize
region F as the factorial Schur function $s_\nu(x_1,\ldots,x_k|y_1,\ldots,y_n)$,
where $\nu$ is the Young diagram, read from bottom to top, that encodes the green lines at the boundary
between F and J. And J is precisely a MS-puzzle (cf section \ref{puzzles}), which contributes
$e^\nu_{\lambda,\mu}(y_1,\ldots,y_n;z_1,\ldots,z_n)$.
So the right hand side equals
$\sum_\nu e^\nu_{\lambda,\mu}(y_1,\ldots,y_n;z_1,\ldots,z_n)s_\nu(x_1,\ldots,x_k|y_1,\ldots,y_n)$.

Finally, we find the desired equality
\begin{multline*}
s_\lambda(x_1,\ldots,x_k|z_1,\ldots,z_n)s_\mu(x_1,\ldots,x_k|y_1,\ldots,y_n)
\\=
\sum_\nu e^\nu_{\lambda,\mu}(y_1,\ldots,y_n;z_1,\ldots,z_n)s_\nu(x_1,\ldots,x_k|y_1,\ldots,y_n)
\end{multline*}
The summation over $\nu$ is only on Young diagrams inside the rectangle $k\times(n-k)$, which is related to the fact
that (i) we have only $k$ variables $x_i$ and (ii) the sum of widths of $\mu$ and $\lambda$ satisfies $\mu_1+\lambda_1\le n-k$.
%
If one sets all $y_i$ and $z_i$ to zero, then 
the equality becomes
\[
s_\lambda(x_1,\ldots,x_k)s_\mu(x_1,\ldots,x_k)=\sum_\nu c^\nu_{\lambda,\mu}s_\nu(x_1,\ldots,x_k)
\]
which is the product formula characterizing the usual Littlewood--Richardson coefficients.

\subsection{Alternate solution of the Molev--Sagan problem}\label{altms}
\begin{figure}
\psfrag{k}[Bc]{$\overbrace{\hbox to 1.6cm{\hfill}}^k$}%
\psfrag{n}[cc][][1][63]{$\underbrace{\hbox to 2.6cm{\hfill}}$}\psfrag*{n}[lc]{\quad$n$}%
\psfrag{n2}[cc][][1][117]{$\overbrace{\hbox to 2.6cm{\hfill}}$}\psfrag*{n2}[lc]{$\!\!\!\!\!n$}%
\psfrag{=}[][][1.5]{$=$}\psfrag{theta}{$\lambda$}\psfrag{mu}{$\mu$}%
\psfrag{y1}{$\scriptstyle y_1$}\psfrag{y2}{$\scriptstyle y_2$}\psfrag{y3}{$\scriptstyle y_3$}\psfrag{y4}{$\scriptstyle y_4$}\psfrag{y5}{$\scriptstyle y_5$}%
\psfrag{z1}{$\scriptstyle z_1$}\psfrag{z2}{$\scriptstyle z_2$}\psfrag{z3}{$\scriptstyle z_3$}\psfrag{z4}{$\scriptstyle z_4$}\psfrag{z5}{$\scriptstyle z_5$}%
\psfrag{x1}{$\scriptstyle x_1$}\psfrag{x2}{$\scriptstyle x_2$}\psfrag{x3}{$\scriptstyle x_3$}%
\myincludegraphics{0.5}{1.15}{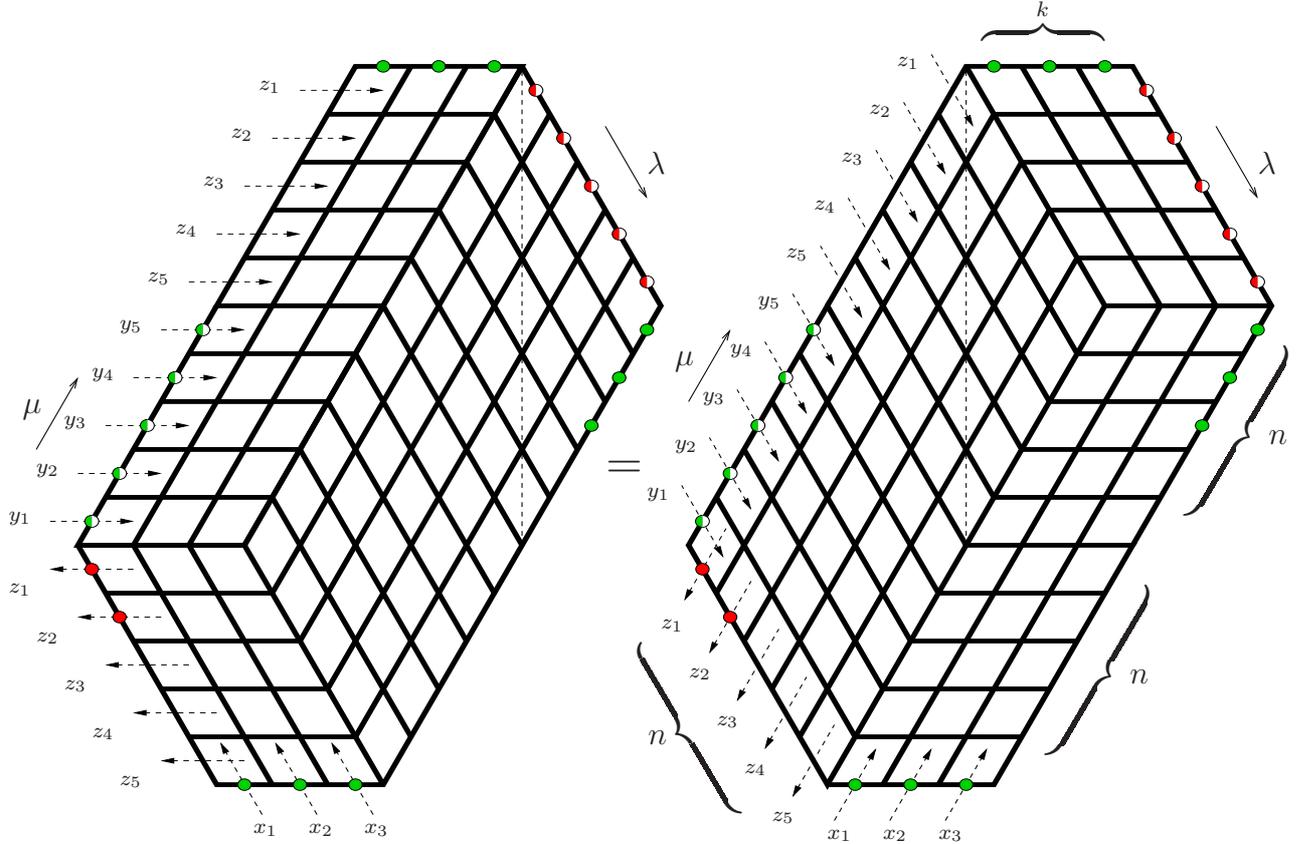}%
\caption{Alternate graphical solution of the MS problem ($n=5$, $k=3$).}
\label{fighexams2}
\end{figure}
\begin{figure}
\psfrag{theta2}{$\smallboxes\mskip-40mu\lambda=\tableau{&\\ \\}$}\psfrag{mu2}{$\smallboxes\mskip-45mu\mu=\tableau{\\ \\}$}
\psfrag{theta}{$\lambda$}\psfrag{mu}{$\mu$}\psfrag{nu}{$\nu$}%
\psfrag{A}{A}\psfrag{B}{B}\psfrag{C}{C}\psfrag{D}{\color{blue} D}\psfrag{E}{\color{blue} E}\psfrag{F}{\color{blue} F}\psfrag{G}{G}\psfrag{H}{H}\psfrag{I}{I}\psfrag{J}{\color{blue} J}%
\myincludegraphics{0.5}{1.15}{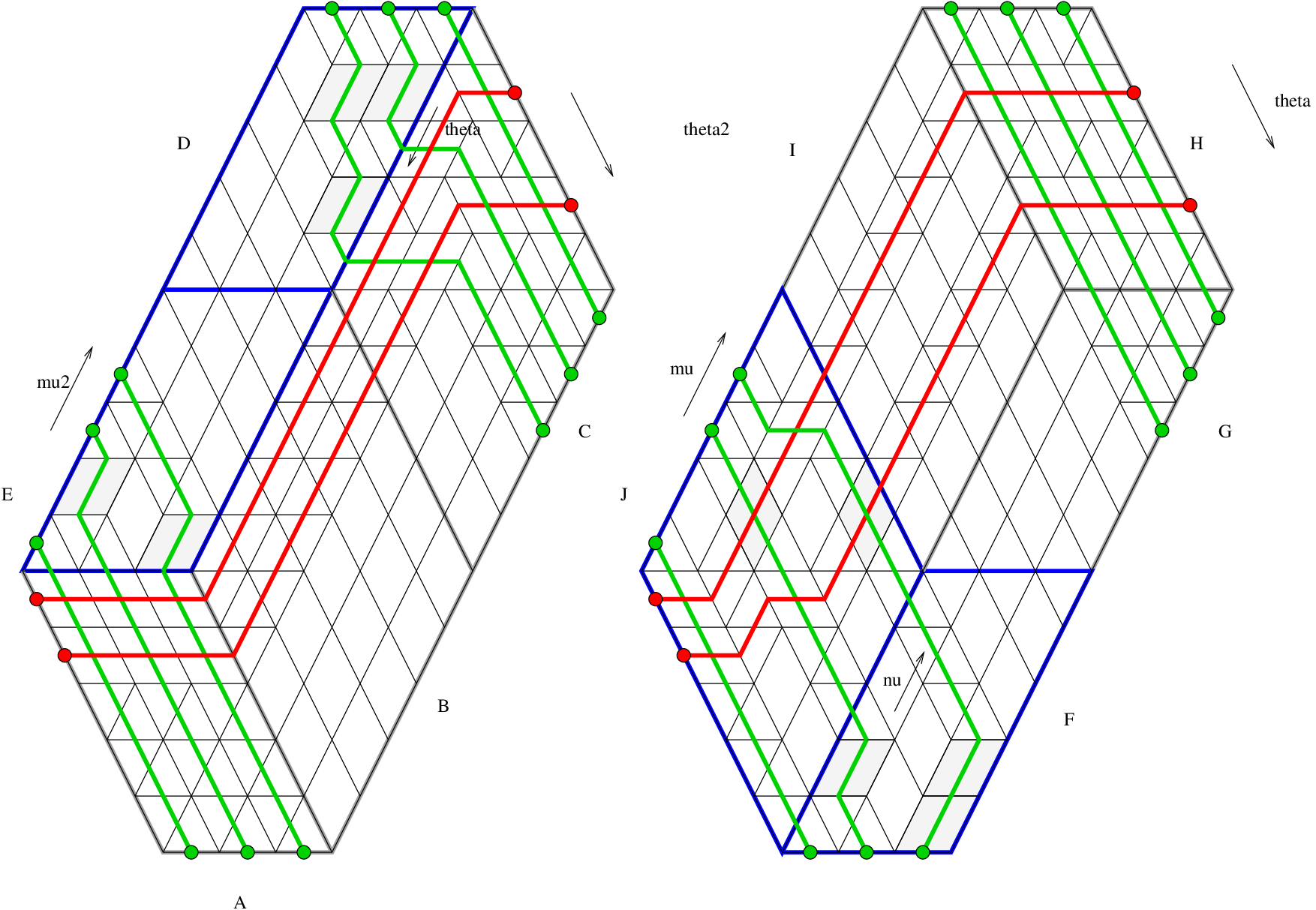}%
\caption{An example of tilings contributing to each side of the equality of Fig.~\ref{fighexams2}.}
\label{fighexams2ex}
\end{figure}
Interestingly, there is a small variation of the construction of the previous section, which produces another, not obviously equivalent, solution
of the Molev--Sagan problem. Since it is based on the same principle, we shall only sketch the derivation, based on Figs.~\ref{fighexams2} and \ref{fighexams2ex}.

By inspection, one easily finds that regions A, B, G, H, I are frozen, with lines going straight as indicated
on the figure. Region C is strictly speaking not frozen; in fact, we recognize an upside-down (180 degree rotated) MS-puzzle,
so that its weight is equal to $e^{\lambda'}_{\lambda,\varnothing}(z_1,\ldots,z_n;z_1,\ldots,z_n)$ where $\lambda'$
is the upper left side of C read top to bottom. This, according to the previous section, is the coefficient
of $s_{\lambda'}(x_1,\ldots,x_k|z_1,\ldots,z_n)$ in the expansion of $s_{\lambda}(x_1,\ldots,x_k|z_1,\ldots,z_n)\times 1$ (note the equality of second alphabets!). So it is equal to $\delta_{\lambda}^{\lambda'}$ (and on the picture
we have shown the unique $\lambda=\lambda'$ configuration).

The other regions are treated the same way as before. Regions D and E correspond to
$s_\lambda(x_1,\ldots,x_k|z_1,\ldots,z_n)$ and $s_\mu(x_1,\ldots,x_k|y_1,\ldots,y_n)$.
If $\nu$ encodes as before the edges between F and J,
then region F corresponds to
$s_\nu(x_1,\ldots,x_k|y_1,\ldots,y_n)$, 
on condition that $\lambda_1+\mu_1\le n-k$. 
Finally region J is an upside-down MS-puzzle with the $y_i$ in the reverse order, so that
we find the equality
\begin{multline*}
s_\lambda(x_1,\ldots,x_k|z_1,\ldots,z_n)s_\mu(x_1,\ldots,x_k|y_1,\ldots,y_n)
\\=
\sum_\nu e^{\bar\mu}_{\lambda,\bar\nu}(y_n,\ldots,y_1;z_1,\ldots,z_n)s_\nu(x_1,\ldots,x_k|y_1,\ldots,y_n)
\end{multline*}

As a corollary, we find the curious identity:
\[
e^\nu_{\lambda,\mu}(y_1,\ldots,y_n;z_1,\ldots,z_n)=
e^{\bar\mu}_{\lambda,\bar\nu}(y_n,\ldots,y_1;z_1,\ldots,z_n)
\]

\subsection{The Knutson--Tao problem}
There is a more basic problem which consists in expanding the product of two factorial Schur functions with the same {\em two}
sets of variables as a sum of factorial Schur functions, and more specifically
providing a manifestly positive formulas for the structure coefficients
in the sense of Graham.
It was solved in various ways \cite{K-ELR,M-ELR,KT}, but here we are
particularly interested in the solution in terms of puzzles,
as in \cite{KT}; this we call the Knutson--Tao (KT) problem.

Clearly, a solution of the MS problem provides a solution of the KT problem by setting $y_i=z_i$, $i=1,\ldots,n$.
Thus, the previous two sections provide two solutions of the KT problem, which turn out to be different. 
The most interesting one is the alternate one:
when $y_i=z_i$, the upper half of the rhombus J of the right hand side
of Fig.~\ref{fighexams2ex} becomes an equivariant puzzle
(as in the example shown), with the spectral parameters labelled in the proper order,
and the equality of section \ref{altms} becomes
\begin{equation*}
s_\lambda(x_1,\ldots,x_k|z_1,\ldots,z_n)s_\mu(x_1,\ldots,x_k|z_1,\ldots,z_n)
=
\sum_\nu c^{\nu}_{\mu,\lambda}(z_1,\ldots,z_n)s_\nu(x_1,\ldots,x_k|z_1,\ldots,z_n)
\end{equation*}
by definition of the $c^{\nu}_{\mu,\lambda}(z_1,\ldots,z_n)$, cf section \ref{puzzles}. This is precisely the formula found in \cite{KT}.

\appendix

\section{Square-triangle-rhombus tilings}
As mentioned in the introduction, it is known that the tiling model introduced in section \ref{tilingmodel}
is related to the square-triangle tiling model. More precisely, the latter is equivalent to the tiling
model without the shaded tiles $\gamma$. We describe here the slightly more general (square-triangle-rhombus) 
tiling model which includes
the shaded tiles (in the spirit of \cite{dGN-sqtriunpub}). Though it is not needed anywhere in this paper,
the connection is worth mentioning.

\begin{figure}
\psfrag{a}{$\alpha_-$}\psfrag{b}{$\alpha_+$}\psfrag{c}{$\beta_-$}\psfrag{d}{$\beta_+$}\psfrag{e}{$\beta_0$}\psfrag{f}{$\gamma_-$}\psfrag{g}{$\gamma_+$}\psfrag{h}{$\gamma_0$}
\includegraphics[scale=0.5]{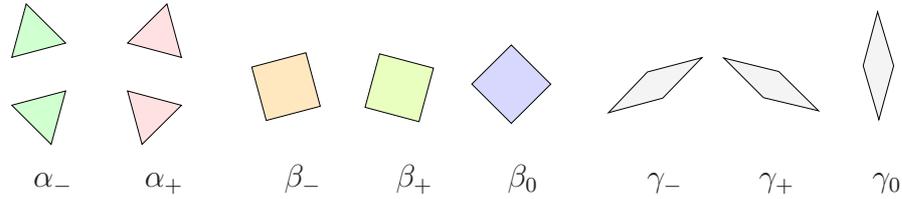}
\caption{Tiles of the square-triangle-rhombus tiling model. The colors are purely decorative.}
\label{figdeformedtiles}
\end{figure}

Consider tiles of three types: equilateral triangles, squares and ``thin rhombi'' with angles 30 and
150 degrees. All of them have sides of unit length. The square-triangle-rhombus tiling model consists
in filling a region of the plane with these tiles, with an addition restriction on the allowed rotations
of the thin rhombi which is the following.
It is easy to see that all edges can occur in exactly six directions which differ from each other by
multiples of 30 degrees. We may for example assume that they are of the form $15+30k$ degrees, $k$ integer.
Thus, there are four possible rotations of triangles, three rotations of square, and six rotations
of thin rhombi among which we select only three as shown on Fig.~\ref{figdeformedtiles}.

Now consider the following operation on a given tiling configuration: deform it in such
a way that the direction of every edge can only take three values, which is the closest slope of the form
$60k$, $k$ integer. One can show that this can be done consistently (as always
in the theory of random and quasi-periodic tilings, these tilings can be seen as projections of
a two-dimensional surface embedded in a higher dimensional space -- here, four -- and the transformation
corresponds to changing the direction of projection); furthermore, mark with a $\pm$
the edge which have been rotated $\pm 15$ degrees counterclockwise. In this way the triangles
become one of the four tiles $\alpha$ of Fig.~\ref{figtiles}; 
similarly, squares become pairs of tiles of type $\beta$;
and thin rhombi become pairs of tiles $\gamma$. The operation is invertible
and is thus the desired equivalence.
As before, if we want to preserve integrability,
we must allow only one of the three types of thin rhombi in any given region.

\begin{figure}
\psfrag{+}{$\scriptscriptstyle +$}\psfrag{-}{$\scriptscriptstyle -$}
\psfrag{a}{(a)}\psfrag{b}{(b)}
\includegraphics[scale=0.6]{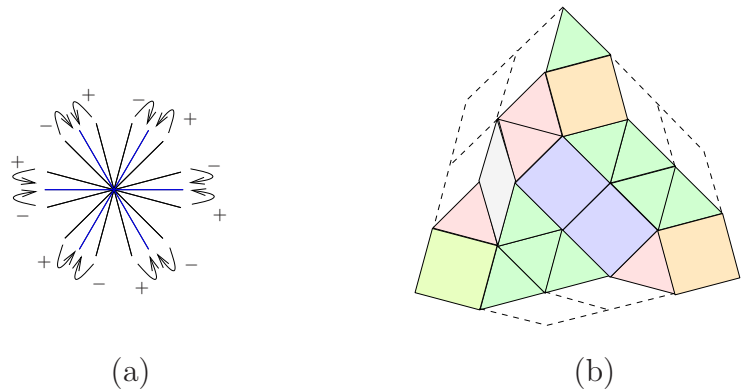}
\caption{(a) The 6 directions of the generalized square-triangle model and how to deform them
into the 3 directions (in blue) of the regular triangular lattice. (b) Example of square-triangle-rhombus tiling corresponding to an equivariant puzzle.}
\label{figsquaretriangle}
\end{figure}

For example, starting from Fig.~\ref{figsquaretriangle} one
obtains the equivariant puzzle of the right of Fig.~\ref{figequivpuzzle}. 
Note that one can read directly the three Young diagrams encoded in the three sides of the triangle
(as was observed in \cite{Pur} in the non-equivariant case), by completing the deformed puzzle
into a hexagon with sides of lengths $k,n-k,k,n-k,k,n-k$ and drawing the complement of the puzzle as
sets of the three ``forbidden'' rotations of the thin rhombi.
One must be careful that 
one recovers this way the three Young diagrams corresponding
to the binary strings read {\em clockwise}.

Another interesting observation is that the use of the Yang--Baxter equation (proposition \ref{ybe}) corresponds in this new
language to two elementary operations on tilings (cases (i,ii) and (iii) in the proof),
up to reflection and rotation:
\[
\vcenter{\hbox{\psfrag{=0}{$=$}
\includegraphics[scale=0.6]{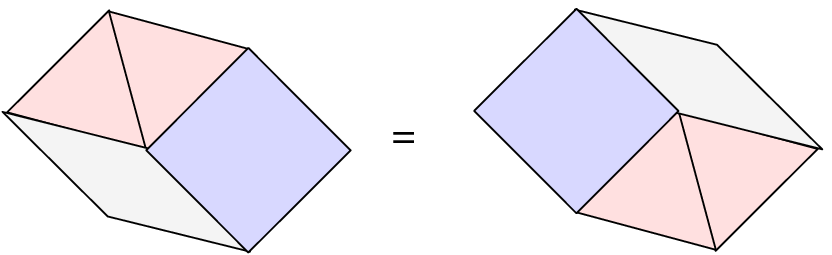}}}
\quad \text{and}\quad
\vcenter{\hbox{\psfrag{+}{$+$}\psfrag{=0}{$=\ 0$}
\includegraphics[scale=0.6]{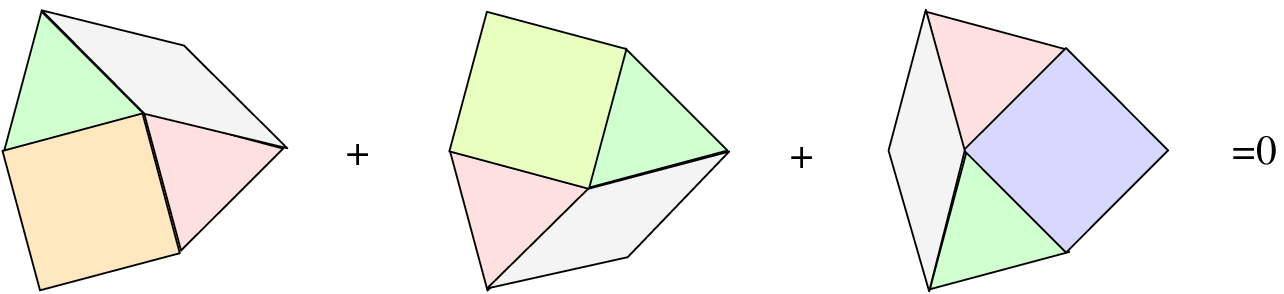}}}
\]
Note the strong similarity with the migration of \cite{Pur}; 
the latter is roughly the special case of the Yang--Baxter
equation where one spectral parameter is set to zero.

\renewcommand\MR[1]{\relax\ifhmode\unskip\spacefactor3000 \space\fi
  \MRhref{#1}{{\sc mr}}}
\renewcommand{\MRhref}[2]{%
  \href{http://www.ams.org/mathscinet-getitem?mr=#1}{#2}}

\bibliography{../biblio}
\bibliographystyle{amsplainhyper}

\end{document}